\documentclass[%
 reprint,
superscriptaddress,
 amsmath,amssymb,
 aps,
prb,
floatfix
]{revtex4-2}
\usepackage{chemformula} 
\usepackage[T1]{fontenc} 

\usepackage{graphicx}    
\usepackage{mhchem}
\usepackage{caption}
\usepackage{subcaption}  
\usepackage{siunitx}
\usepackage{gensymb}

\begin{document}
\author{Michiel P.F. Wopereis}
\affiliation{Department of Precision and Microsystems Engineering, Delft University of Technology,
Mekelweg 2, 2628 CD Delft, The Netherlands}

\author{Niels Bouman}
\affiliation{Department of Precision and Microsystems Engineering, Delft University of Technology,
Mekelweg 2, 2628 CD Delft, The Netherlands}

\author{Satadal Dutta}
\affiliation{Department of Precision and Microsystems Engineering, Delft University of Technology,
Mekelweg 2, 2628 CD Delft, The Netherlands}

\author{Peter G. Steeneken}
\affiliation{Department of Precision and Microsystems Engineering, Delft University of Technology,
Mekelweg 2, 2628 CD Delft, The Netherlands}
\author{Farbod Alijani}
\affiliation{Department of Precision and Microsystems Engineering, Delft University of Technology,
Mekelweg 2, 2628 CD Delft, The Netherlands}
\author{Gerard J. Verbiest}
\affiliation{Department of Precision and Microsystems Engineering, Delft University of Technology,
Mekelweg 2, 2628 CD Delft, The Netherlands}
\email{G.J.Verbiest@tudelft.nl}






\title[Tuning dissipation dilution in 2D material resonators by MEMS-induced tension]{Tuning dissipation dilution in 2D material resonators by MEMS-induced tension}


\begin{abstract} 
Resonators based on two-dimensional (2D) materials have exceptional properties for application as nanomechanical sensors, which allows them to operate at high frequencies with high sensitivity. However, their performance as nanomechanical sensors is currently limited by their low quality ($Q$)-factor. Here, we make use of micro-electromechanical systems (MEMS) to apply pure in-plane mechanical strain, enhancing both their resonance frequency and Q-factor. In contrast to earlier work, the 2D material resonators are fabricated on the MEMS actuators without any wet processing steps, using a dry-transfer method. A platinum clamp, that is deposited by electron beam-induced deposition, is shown to be effective in fixing the 2D membrane to the MEMS and preventing slippage. By in-plane straining the membranes in a purely mechanical fashion, we increase the tensile energy, thereby diluting dissipation. This way, we show how dissipation dilution can increase the $Q$-factor of 2D material resonators by 91\%. The presented MEMS actuated dissipation dilution method does not only pave the way towards higher $Q$-factors in resonators based on 2D materials, but also provides a route toward studies of the intrinsic loss mechanisms of 2D materials in the monolayer limit. 
\end{abstract}

\maketitle
\section{Introduction}

Nanomechanical resonators made of two-dimensional (2D) materials are the subject of intensive research due to their remarkable properties. Their low mass, combined with their high Young's modulus, leads to resonance frequencies that are typically a few tens of MHz \cite{Sakhaee-Pour2008-RN84}. Yet, their extreme flexibility in the out-of-plane direction enhances the sensitivity to external stimuli and makes them promising for various applications, including mass \cite{Lassagne2008-RN44, Sakhaee-Pour2008-RN84, Atalaya2010-RN94, Singh2010-RN4}, force\cite{Weber2016-RN20, Lemme2020-RN3, Roslon2022-RN157},  pressure\cite{Lemme2020-RN3, RomijnYear-RN7, Dolleman2016-RN88, Shahdeo2020-RN5, Choi2020-RN16}, and temperature sensing\cite{Lemme2020-RN3, Dolleman2020-RN99, Hanqing2023-RN180}.

The performance of nanomechanical resonant sensors and clocks is generally limited by their dissipation per cycle ($\Delta W$). A low $\Delta W$ results in a high quality ($Q$-)factor, which is the ratio of stored energy $W$ to $\Delta W$ over a single oscillation cycle ($Q = 2\pi W /\Delta W$). A low $\Delta W$ and thus high $Q$-factor physically insulates the resonator from external noise sources, allowing long-term coherent oscillations while minimizing energy dissipation to the environment \cite{Sementilli2022-RN191, Miller2018-RN128}. Thus, enabling low phase-noise oscillators \cite{Leeson1966-RN213} and high-performance noise-rejection filters \cite{Miller2018-RN128, Gabrielson1995-RN210, Gabrielson1993-RN211}.



Increasing the tensile stress can be an effective strategy to realize high-$Q$  resonators \cite{Schmid2016-RN203, Verbridge2006-RN200}. The tension leads to an increase in the stored energy $W$ without significantly affecting losses, thereby increasing the ratio $W/\Delta W$ and the $Q$-factor \cite{Fedorov2019-RN218}, an effect that is commonly known as dissipation dilution. Since this strategy was very successful in realizing high-$Q$ resonators in SiN that can be grown with high intrinsic tensile stress \cite{Verbridge2006-RN200}, it was also considered as a method for increasing the quality factor of 2D materials that could not be grown with high intrinsic stress, such that they might become serious contenders for high-$Q$ sensors and high $fQ$ resonant quantum devices. Out-of-plane electrostatic and thermal forces were used to achieve this quality factor tuning \cite{Davidovikj2018-RN96}. However, attempts to increase the $Q$-factor through out-of-plane electrostatic gating of 2D material membranes typically result in a reduction of the $Q$-factor instead of an increase by dissipation dilution\cite{Weber2016-RN20, Singh2010-RN4, Morell2016-RN228}. This reduction is due to the voltage-dependent electronic Joule dissipation of the displacement current within the resonator \cite{Morell2016-RN228, Song2012-RN140}. Furthermore, the out-of-plane electrostatic pulling force increases side wall adhesion, which facilitates dissipation through coupling with the substrate \cite{Rieger2014-RN229, Davidovikj2018-RN96, Steeneken2021-RN1}. Thermal expansion-based tuning strategies to increase $Q$ \cite{Davidovikj2018-RN96} have the drawback of making it difficult to distinguish tension effects from other thermal effects on $Q$. For example, the change in membrane temperature changes material parameters that can increase damping via, e.g., the thermoelastic dissipation mechanism \cite{Liftshitz2000-RN193, Miller2018-RN128}.

Considering these drawbacks, it would be ideal to apply pure in-plane mechanical stress to increase the $Q$ of 2D resonators. To reach this goal, special micro-electromechanical system (MEMS) actuators were developed in earlier works \cite{Verbiest2016-RN11,Perez-Garza2014-RN9,Xie2021-RN59}. However, all these approaches have a step in which liquids such as glues or etchants are in contact with the 2D material thereby affecting the device properties and $Q$. Moreover, the involved liquids and complex fabrication process results in a low device yield making it difficult to show the clamping efficiency of 2D materials on MEMS actuators. The field is thus in need of a simplified, dry fabrication method for MEMS actuators with integrated and clamped 2D materials.
In this paper, we achieve this goal and provide evidence of dissipation dilution in suspended 2D material resonators that are controllably tensioned in-plane using a micro-electromechanical systems (MEMS) actuator. To enable these experiments, we introduce a dry-transfer method to precisely suspend 2D materials over MEMS gaps and rigidly clamp them with a layer of platinum using electron-beam-induced deposition (EBID). We actuate the membrane resonances optothermally and record the resulting motion using an interferometry setup, from which we extract the $Q$-factor ($Q$) and resonance frequency ($f_0$). By applying strain with the MEMS actuator to the 2D material resonator in a mechanical and controllable fashion, we find an increase in resonance frequency as well as the $Q$-factor, which are consistent with dissipation dilution. Our findings thus provide a new way for enhancing the Q-factor via dissipation dilution in 2D material resonators realized with a dry fabrication method. 


\section{Fabrication}
\label{sec:fabrication}
         
\begin{figure*}[htp]
    \centering
    \includegraphics{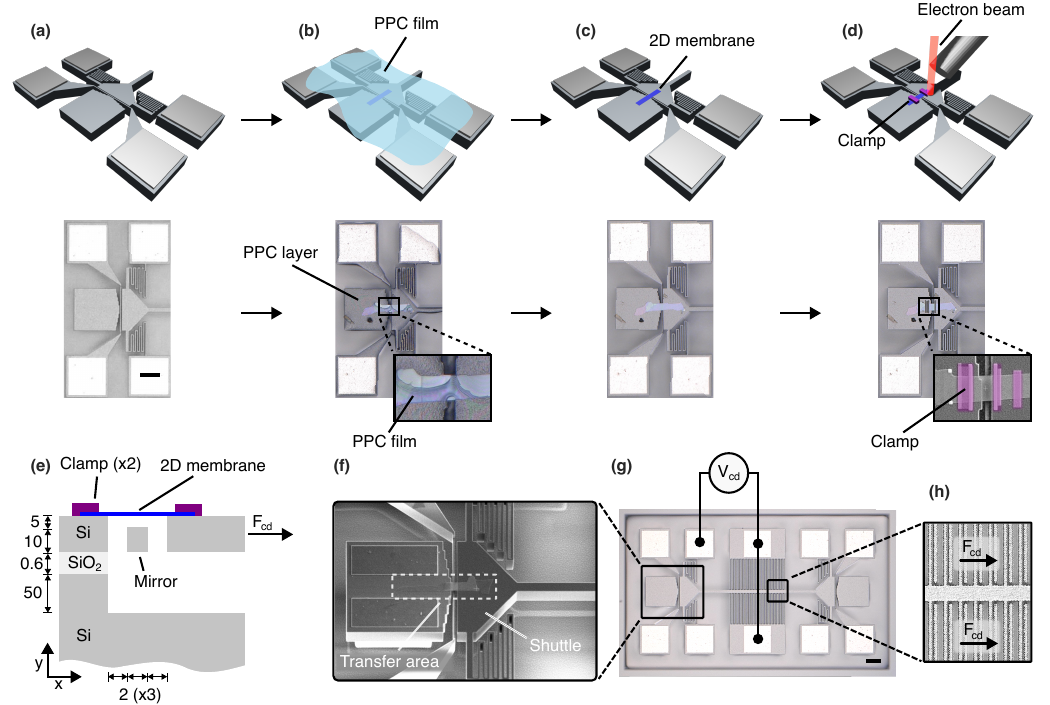}
    \caption{Device geometry: drawings, optical and SEM micrographs. a) Device straining platform. Scale bar: 50 micrometers. b) Device after the membrane transfer using a sacrificial PPC layer. c) Device after annealing. d) Illustration of the EBID process, an optical image of the device after clamping, and a false-colored SEM image highlighting the platinum clamps. e) Side view of the final device. Dimensions are in micrometers. f) SEM image under an angle of the device. g) Top view of an empty MEMS actuator. Scale bar: 50 micrometers. Voltage over comb drive fingers ($V_\mathrm{cd}$). h) Detailed view of the comb drive indicating the direction of the comb drive force ($F_\mathrm{cd}$).  
    \label{fig:device-characterization} }
\end{figure*}

We use a MEMS actuator that is designed and fabricated in the commercially available XMB10 process from X-FAB \cite{ZouYear-RN231, StreitYear-RN230}. The resulting device (see Fig.~\ref{fig:device-characterization}) consists of a moving shuttle with 38 comb fingers that is held suspended by four serpentine flexures. The flexures are connected to fixed anchors that have aluminum bond pads for making electrical contact by wire bonding. The crystalline silicon shuttle has a thickness of 15 $\mu$m, length of 520 $\mu$m, comb finger length of 103 $\mu$m, and asymmetric finger spacing of 2.0 $\mu$m and 4.0 $\mu$m. The membranes are suspended over a 6 \textmu m trench between the fixed anchor and the moving shuttle. Inside the trench, at a depth of 5 $\mu$m below the shuttle surface, a suspended silicon beam acts as a mirror for interferometric readout of the membrane motion. In order to transfer the membranes onto the MEMS actuator, we first mechanically exfoliate 2D materials \cite{Novoselov2005-RN222} onto a 5 mm $\times$ 5 mm PDMS sheet on a microscope slide. Next, we use a microscope to select membranes on the PDMS sheet with a minimum length of 20 \textmu m such that they can cover the suspended trench as well as parts of the fixed anchor and the moving shuttle. Membranes are selected based on their flatness and uniformity. Once we find a suitable membrane, we use a dome-shaped PDMS stamp covered with a sacrificial polypropylene carbonate (PPC) film to pick it up from the PDMS sheet\cite{Kinoshita2019-RN41}. The utilization of a PDMS dome results in a smaller contact area with the MEMS actuator, approximately 350 \textmu m in diameter, which allows the precise positioning of a membrane while minimizing contaminations. 
We then bring the membrane on the PDMS dome in contact with the MEMS actuator and heat the stage to 110$\degree $C (above the melting point of PPC). This causes the PPC film to melt and ensures the transfer of the membrane onto the designated area (see Fig.~\ref{fig:device-characterization}(b)). After the transfer, we wire bond the MEMS actuator and connect all terminals to a common ground; this crucial step prevents any electrostatic force-induced movement during the rest of the fabrication process. Next, we remove the PPC film from the membrane and the MEMS actuator through annealing in a high vacuum oven at a pressure below $10^{-5}$ mbar for 3 hours at a temperature of 300 $\degree $C (exceeding the decomposition temperature of PPC\cite{Wang1997-RN189}). After annealing, we inspect the sample optically to confirm the PPC removal (see Fig.~\ref{fig:device-characterization}(c)). 
Finally, we clamp the membrane with a layer of platinum using EBID \cite{Lee2019-RN60}. Detailed information on the fabrication procedure is available in Supporting Information S1.


In total, we fabricated 4 devices (D1-D4) with different 2D materials using the method outlined above (see Table \ref{tab:fabricated-devices}). Figure~\ref{fig:device-characterization}(e) shows a schematic cross-section of a device including the dimensions. All membranes have a suspended length of 6\textmu m, width $w$, and thickness $t$, as determined with a white light interferometer (see Table~\ref{tab:fabricated-devices}).

To mechanically tension 2D materials with the MEMS actuators (see Figs.~\ref{fig:device-characterization}(f)-(g)), we apply a potential difference $V_\mathrm{cd}$ between the comb fingers. Due to the asymmetric placement of the comb fingers (see Fig.~\ref{fig:device-characterization}(h)), a force $F_\mathrm{cd} = -\tfrac{1}{2}\tfrac{\partial C}{\partial x} V_\mathrm{cd}^2$ will act on the moving shuttle and will tension the membrane and the four serpentine flexures. Here, $\tfrac{\partial C}{\partial x}$ is the change in capacitance $C$ between the comb fingers with respect to a change in position $x$ of the moving shuttle.

Based on their geometry, the MEMS actuators used here can controllably strain membranes up to $\approx$ 11\%, which is set by the maximum in-plane displacement of $\approx$ 0.67 \textmu m (1/3 of the 2 $\mu$m actuation gap) over a 6 \textmu m suspended length. As the maximum in-plane displacement is the limiting factor, pre-deformations and wrinkles in 2D materials \cite{Nicholl2015-RN119, Davidovikj2018-RN96, Steeneken2021-RN1} can reduce the maximum achievable strain. Moreover, the bare MEMS actuator has a pull-in voltage of 13.5 $\pm$ 0.5 V (see Supporting Information 2). As a result, if the stiffness of the 2D membrane stiffness is low, e.g. because it is wrinkled, the MEMS actuator will collapse at this voltage. However, for certain devices, the stiffness of the 2D membrane was high, such that it increased the total device stiffness significantly with respect to that of the MEMS springs and, as a consequence, led to an increase in the pull-in voltage. We were able to apply voltages up to 60 V for device D4 without a pull-in, which demonstrates that the force and stiffness provided by the 2D material were substantial.

\begin{table*}[htp]
    \begin{tabular}{llllllll}
        \hline
          Device & Material & Clamped & w ($\mu m$) & t (nm) & $Q_\mathrm{int} / f_\mathrm{plate}^2 $ ($Hz^{-2}$) & $Q_\mathrm{int}$ & $Q_\mathrm{ext}$    \\
         \hline
         D1 & Graphene & No & 7.2  & 9.8 $\pm$ 0.33  & 7.084 $\times\,10^{-12}$   & < 71 & 182 \\
         D2 & Graphene & No & 16.7 & 22.47 $\pm$ 0.15  & 2.106 $\times\,10^{-12}$ & < 127 &> 1000 \\
         D3 & MoS2 & Yes & 19.9 & 77.79$ \pm$ 0.33    & 1.362 $\times\,10^{-12}$  & < 64 &180 \\
         D4 & WS2 & Yes & 16.0  & 94.39 $\pm$ 0.23    & 2.515 $\times\,10^{-12}$  & < 203 &> 1000 \\
    \end{tabular}
    \caption{Characteristics of fabricated devices D1-D4, including 2D material, clamped using EBID of platinum, width $w$, thickness $t$ and fitted parameters $Q_\mathrm{int}/f_\mathrm{plate}^2$ and $Q_\mathrm{ext}$, where, $Q_\mathrm{int}$ is the intrinsic quality factor due to bending and elongation losses, $f_\mathrm{plate}$ is the resonance frequency due to the bending rigidity and $Q_\mathrm{ext}$ is damping from extrinsic dissipation sources (see Eq.~\eqref{eqn:Q-approximation}). The membrane thickness is determined using AFM. The width is obtained by measuring an optical or SEM image. }
    \label{tab:fabricated-devices}
\end{table*}


\section{Measurements}
\label{sec:results}
\begin{figure*}[htp]
    \centering
    \includegraphics{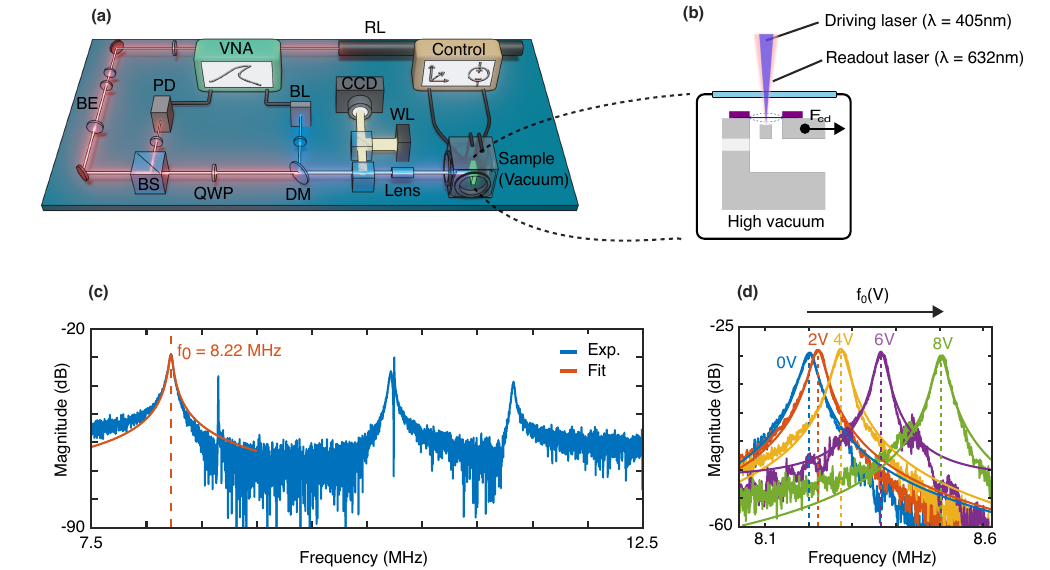}
    \caption{Device measurements. (a) Fabry-Perot Interferometry setup for measuring the resonance frequencies. Vector Network Analyser (VNA), $V_\mathrm{cd}$ and stage control (control), Beam expander (BE), Beam splitter (BS), Quarter wave plate (QWP), Dichroic mirror (DM), White light source (WL), Photodetector (PD), Red ($\lambda=632$ nm) He-Ne Laser (RL), Blue ($\lambda=405$ nm) Diode Laser (BL). (b) Side view of the device inside the high vacuum chamber. (c) Fitting the harmonic oscillator equation on the experimental data of the VNA for device D4 at $V_\mathrm{cd}=2$ V to extract the resonance frequency ($f_0$) and quality factor (Q). (d) experimental data of the VNA for device D4 for $V_\mathrm{cd}=0$ V, 2 V, 4 V, 6 V, and 8 V (DC).
    \label{fig:measurement-setup}}
\end{figure*}

We measure the dynamics of the membranes using an optical interferometry setup (see Fig.~\ref{fig:measurement-setup}(a)). The devices are placed in a vacuum chamber with a pressure below $10^{-5}$ mbar and are actuated using a blue diode laser ($\lambda=405$ nm) that is power-modulated through a vector network analyzer (VNA). We use a red He-Ne laser ($\lambda=632$ nm) to measure the motion of the membrane as its reflected intensity highly depends on the distance between the membrane and the mirror (see Fig.~\ref{fig:measurement-setup}(b)). The intensity of the reflected red laser light is detected using a photodiode and further processed by the VNA. Figure~\ref{fig:measurement-setup}(c) shows a typical response recorded by the VNA. We observe multiple peaks in the spectrum that we identify as resonance frequencies. To analyze these further, we fit the frequency response $M$ of the fundamental resonance to the well-known harmonic oscillator model\cite{Schmidt2020-RN196, Figliola2014-RN197, Miller2018-RN128}:


\begin{equation}
\label{eqn:harmonic-oscillator-fit}
M(f) = \frac{\left(A f^2 / Q\right)}{\sqrt{\left(f_0^2-f^2\right)^2+\left(\frac{f_0 f}{Q}\right)^2}},
\end{equation}

\noindent where $f_0$ is the resonance frequency in Hz, $Q$ is the quality factor, and $A$ is the peak amplitude. Figure \ref{fig:measurement-setup}(c) shows this fit on the frequency response of device D4 at $V_\mathrm{cd} = 2$ V, from which we extract a resonance frequency of 8.22 MHz. We then repeat the measurement for different $V_\mathrm{cd}$ (see Fig.~\ref{fig:measurement-setup}(d)) to analyze the dependence of $f_0$ and $Q$ on the applied force $F_\mathrm{cd}$. As Fig.~\ref{fig:measurement-setup}(d) shows, we see an increase in $f_0$ with applied $V_\mathrm{cd}$ due to the increase in tension in the membrane, which is in agreement with results reported in the literature \cite{Verbiest2021-RN15, Goldsche2018-RN40, SonntagYear-RN122, Xie2021-RN59}.


\section{Clamping of 2D material resonators}
\begin{figure}[htp]
    \centering
    \includegraphics{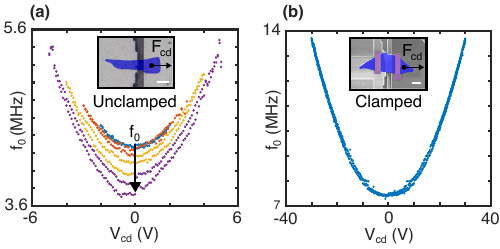}
    \caption{Comparison of the resonance frequency ($f_0$) versus the comb drive voltage ($V_\mathrm{cd}$). $F_\mathrm{cd}$ indicates the pulling direction of the suspended shuttle by the comb drive actuator. Scale bars: 10 micrometers. a) Unclamped device D1, $V_\mathrm{max}$ = 2 V (blue), 3 V (orange), 4 V (yellow) and 5V (purple), false-colored optical image (blue: 2D membrane) b) Device D3 clamped with a platinum layer deposited by EBID,$V_\mathrm{max}$ = 30 V (blue), false-colored SEM image (blue: 2D membrane, purple: platinum clamps)}
    \label{fig:clamping-effectiveness}
\end{figure}

To evaluate the effectiveness of a deposited layer of platinum using EBID in clamping 2D material resonators, we compare the response of an unclamped device (D1) to the response of a clamped one (D3). We vary the comb-drive force $F_\mathrm{cd}$ for both devices by varying $V_\mathrm{cd}$. In a single sweep, we start at $V_\mathrm{cd} = 0$ V, increase the voltage to a maximum $V_\mathrm{max}$, return to 0 V, decrease the voltage to $-V_\mathrm{max}$, and finally return to 0 V. For each subsequent sweep, we increase $V_\mathrm{max}$ by 1 V up to a maximum of 5 V for device D1, to gradually increase the maximum force exerted on the membrane and monitor the shift in the fundamental resonance. Figure~\ref{fig:clamping-effectiveness} shows $f_0$ as a function of $V_\mathrm{cd}$ extracted from one such experiment for devices D1 and D3. For the unclamped device (see Fig.~\ref{fig:clamping-effectiveness}(a)), we observe a significant decrease in $f_0$ at $V_\mathrm{cd} = 0$ V after each voltage sweep. At the last measurement point at 0 V, $f_0$ shows an irreversible reduction from 4.30 MHz to 3.73 MHz, which corresponds to a decrease of approximately 13\% with respect to the very first measurement on this membrane. The permanent reduction of $f_0$ might be attributed to irreversible slippage, unsticking, or ironing out the wrinkles that increase the effective length of the membrane, reduce its tension, and thus resonance frequency. This is in contrast to recent observations of a reversible sliding scenario where a closed $f_0$ vs. $V_\mathrm{cd}$ loop is expected \cite{Ying2022-RN53}.




In contrast to the unclamped device D1, device D3 contains a clamped membrane, as depicted in the inset of Fig.~\ref{fig:clamping-effectiveness}(b). As Fig.~\ref{fig:clamping-effectiveness}(b) shows, we observe a notably different $f_0$ vs. $V_\mathrm{cd}$ response when compared to the unclamped device. Initially, $f_0$ measures 7.49 MHz, ascending to 13.67 MHz at 30 V and decreasing back to 7.42 MHz at 0 V, a change of less than 1 \%. In subsequent sweeps, the resonance frequency at $V_\mathrm{cd}=0$V remains stable within a range of 0.02 MHz. These measurements show that the resonance frequency $f_0$ of the device D3 is much more stable than that of the unclamped device, even at substantially higher actuation voltages, resulting in a 36 times larger force ($F_\mathrm{cd} \propto V_\mathrm{cd}^2$) than applied in measurements on device D1. This difference between the clamped and unclamped devices indicates that the deposited layer of platinum using EBID is effective in preventing permanent tension reduction during MEMS actuation. Since the EBID clamps are separated by a few microns from the edge of the trench and thus cannot significantly affect unwrinkling and adhesion mechanisms inside the trench, we also conclude that the irreversible changes in $f_0$ in Fig.~\ref{fig:clamping-effectiveness}(a) are most likely due to slippage of a large part of the flake over the silicon surface.

Furthermore, it's worth noting that the maximum voltage of 30 V significantly surpasses the pull-in voltage $V_\mathrm{PI}$ of the bare MEMS actuator, which was determined to be 13.5 $\pm$ 0.5 V (see Supporting information S2), providing evidence that the clamped membrane generates substantial force on the shuttle, thus preventing its pull-in and collapse. Simultaneously, according to Newton's third law, we conclude that the MEMS actuator effectively pulls and strains the clamped membranes. An equivalent observation for unclamped device D2 and clamped device D4 can be found in Supporting information S4.

\section{Dissipation dilution}
\begin{figure}[htp]
    \centering
    \includegraphics{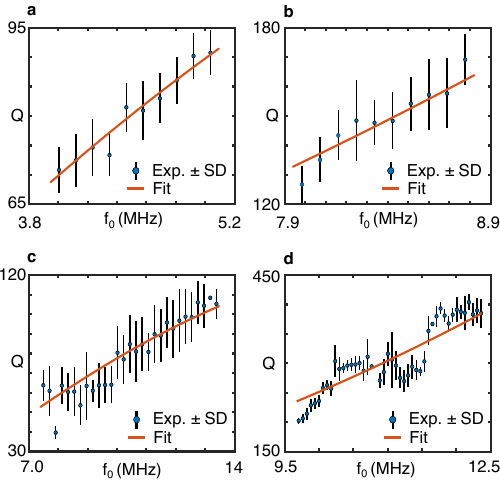}
    \caption{Dissipation dilution in unclamped device D1-D2 and clamped devices D3-D4. Quality factor ($Q$) versus the resonance frequency ($f_0$). The unclamped devices exhibit a decline in $f_0$ and $Q$ due to slippage, in contrast with the clamped devices, which show consistent values for $f_0$ and $Q$ at $V_\mathrm{cd} =$ 0 V. a) Experimental data from unclamped device D1 from the $V_{max} = 4$ V-cycle. The device starts at $f_0 = 4.17$ MHz and $Q =$ 73.5 and stops at $f_0$ = 3.96 MHz and $Q =$ 69.0 due to slippage b) Experimental data from unclamped device D2 from the $V_{max} = 12$ V-cycle. The device starts at $f_0 = 8.45$ MHz and $Q =$ 150 and stops at $f_0$ = 7.95 MHz and $Q =$ 113 due to slippage. c) Experimental data of clamped device D3 from the $V_{max} = 30$ V-cycle. The device maintains a consistent $f_0$ and $Q$ of 7.50 MHz and 63.8, respectively. d) Experimental data of clamped device D4 from the $V_{max} = 60$ V-cycle. The device maintains a consistent $f_0$ and $Q$ of 9.70 MHz and 202, respectively.
   }  
    \label{fig:dissipation-dilution}
\end{figure}

By fitting resonance peaks like in Fig. \ref{fig:measurement-setup}(d), we extract both the quality factor $Q$ and resonance frequency $f_0$. 
To study the effectiveness of the MEMS actuator in tuning the quality factor by dissipation dilution, we plot $Q$ vs $f_0$ for both unclamped and clamped devices over the full actuation voltage range in Fig. \ref{fig:dissipation-dilution}.  It is seen that the $Q$-factor of all devices increases with $f_0$, as expected in a dissipation dilution scenario. For instance, device D4 experiences a 30\% increase in $f_0$ and a 91\% increase in Q. Interestingly, despite the presence of slipping effects, this trend even seems to hold for the unclamped devices D1 and D2 shown in Figs.~\ref{fig:dissipation-dilution}(a) and (b), respectively.


We now compare the experimental relation between $f_0$ and $Q$ with theory. From literature\cite{Sementilli2022-RN191}, the quality factor $Q_D$ of a resonator in the presence of dissipation dilution is given by (see Supporting information S3): 

\begin{equation}
    \label{eqn:Qeb-approximation}
    Q_{D} \approx \left(\frac{f_{0}}{f_\mathrm{plate}}\right)^2 Q_\mathrm{int},
\end{equation}

\noindent where $f_0$ is the measured resonance frequency, $f_\mathrm{plate}$ is the frequency in the presence of bending rigidity, and $Q_\mathrm{int}$ is the intrinsic dissipation. In the presence of other external dissipation mechanisms with quality factor $Q_\mathrm{ext}$, the total $Q$-factor can be further reduced\cite{Miller2018-RN128, Schmid2016-RN203} to a value $Q$:

\begin{equation}
\label{eqn:Q-approximation}
    \frac{1}{Q} = \frac{1}{Q_{D}} + \frac{1}{Q_\mathrm{ext}}
\end{equation}

\noindent We fit Eq.~\eqref{eqn:Q-approximation} to the experimental data in Fig.~\ref{fig:dissipation-dilution} with $Q_\mathrm{int}/f_\mathrm{plate}^2$ and $Q_\mathrm{ext}$ as fit parameters (orange lines). We provide the fitting values in table \ref{tab:fabricated-devices}. Note that the fitted lines are almost linear because they span only a small frequency range.
For the fit parameter $Q_\mathrm{int}/f_\mathrm{plate}^2$, we find a similar order of magnitude for all devices. For devices D1 and D3, we fitted an extrinsic damping contribution of $Q_\mathrm{ext}\,\approx$ 180. For devices D2 and D4, the data could be fitted well without assuming an external dissipation source $Q_\mathrm{ext}$, and therefore, we took $1/Q_\mathrm{ext}=0$. The good agreement between the experimental data and the fitted curves using Eq.~\eqref{eqn:Q-approximation} provides evidence that dissipation dilution can account for the observations. 

\section{Discussion}


In the previous sections, we studied the effect of in-plane stress on the $Q$-factor and resonance frequency of membranes made of multi-layered 2D materials. The results show that the quality factor can be enhanced by generating tension on the 2D material membranes, and the obtained data matches reasonably well with a dissipation dilution model. 
We note that this observation does not rule out other mechanisms that may affect the quality factors. These include (i) mode coupling to other modes of the membrane \cite{Wang2018-RN202, Keskekler2021-RN67, Keskekler2023-RN257}, (ii) a change in dissipation rate $\Delta W$ with applied tension \cite{Zener1937-RN217, Steeneken2021-RN1, Schmid2011-RN194}, (iii) changes in the membrane's geometry with increasing tension, which could be attribued to suppressing wrinkles \cite{Nicholl2015-RN119, Nicholl2017-RN118}, or (iv) the release of edge adhesion during membrane straining \cite{Dolleman2020-RN221}, which facilitates the transfer of energy from the resonating membrane to the substrate, leading to acoustic radiation losses \cite{Steeneken2013-RN216}. Future work could focus on systematically studying these effects as a function of applied tension. 

The main advantage of using a MEMS actuator to apply the tension is that it provides a pure in-plane force, in contrast to thermal, electrostatic gating, or gas pressure-based approaches, where the force is accompanied by other effects that might modify $Q$.
The presented method also has some limitations. First of all, actuator device fabrication, design, and wire bonding are time-consuming. Moreover, the actuation range is limited to 1/3 of the actuator gap, although other actuator designs can provide a more extensive range. Finally, transferring membranes to a movable shuttle is a complicated task and can affect membrane suspending and adhesion to the MEMS actuator. 

A key prospect of the presented work is the ability to apply MEMS tensioning to significantly raise the quality factor of 2D materials, with the aim to eventually scale it down to the monolayer limit and outperform record silicon nitride/carbide devices \cite{Xu2023-RN232, Shin2022-RN242, Cupertino2023-RN243}. Although currently, $Q$s of 2D materials seem to be limited by their low stress and by fabrication artifacts like wrinkles, in the monolayer limit, the intrinsic $Q_\mathrm{int}$ of 2D materials might be much higher than that of multilayer materials since interlayer dissipation mechanisms and thermoelastic damping are largely eliminated \cite{Lindahl2012-RN237}. Moreover, for sensing applications, like resonant mass sensors, the large aspect ratio and low mass of 2D materials can increase sensitivity. 
In this respect, the high ultimate tensile stress of 2D materials like graphene, which was demonstrated to exceed 10 GPa \cite{Goldsche2018-RN14}, can potentially outperform the tension limitations of even silicon carbides \cite{Xu2023-RN232}.

Moreover, it is of interest to evaluate the minimum measured $Q$ in more detail. For device D4, we found a $Q$ of 202. Disregarding the small effect of pre-stress, we expect this $Q$ to be roughly equal to the intrinsic $Q$ since the thickness of the membrane is relatively high and the intrinsic $Q$ is primarily governed by the bending contribution. This intrinsic $Q$ is over 32 times lower than that of silicon nitride, for which a value of $69t~\mathrm{nm}^{-1}$ was found\cite{Villanueva2014-RN215}, which for a thickness equal to that of device D4 of 94.39 nm evaluates to $Q_\mathrm{int}$ = 6513. A question for further study is if the low intrinsic $Q$ in device D4 is due to the intrinsic crystal properties of WS$_2$ compared to SiN or if the value is limited by imperfections such as wrinkles. 
It is clear that the MEMS implementation of dissipation dilution demonstrated in this work, which results in a $Q$ enhancement up to 91\%, is only a first step toward Q-factor engineering of 2D materials and by no means is yet as efficient as that in SiN resonators, where $Q/Q_\mathrm{int}$ factors of over 10$^4$ are reached \cite{Shin2022-RN242}.

\bigskip
\section{Conclusion}
To conclude, we show signatures of dissipation dilution in 2D material resonators. We use a MEMS actuator to strain the membrane uniaxially and thus tune the resonance frequency and $Q$-factor. To induce dissipation dilution, we developed a device fabrication method using dry transfer of membranes on the MEMS actuator and a clamping technique using EBID of platinum to prevent edge-slippage. The MEMS platform can also be used to study slipping and sliding effects, where slipping was observed to reduce both the Q-factor and $f_0$, in line with the dissipation dilution mechanism that increases $Q$ with MEMS actuation force. By pulling on the membranes, we control the in-plane tension of the membranes, which resulted in an increase in resonance frequency $f_0$ of 30\% and an accompanying increase of 91\% in the $Q$ factor. Our results enable a leap in developing higher-$Q$ resonators based on 2D materials with potential applications in sensing, time-keeping, and information processing.
\section*{Supporting Information}
\begin{appendix} 


The following files are available free of charge.
\begin{itemize}
  \item supporting-information.pdf: \\
  S1: Detailed fabrication instructions\\
  S2: MEMS stiffness characterization\\
  S3: Derivation of the dissipation dilution model\\
  S4: $f_0$ vs V measurements for devices D2 and D4
\end{itemize}

\end{appendix} 

\section{Acknowledgements}
We would like to thank the Plantenna research program funded by the 4TU federation and Europractice for providing the design tools and multi-project wafer service.

\bibliography{clamping}

\begin{thebibliography}{68}%
\makeatletter
\providecommand \@ifxundefined [1]{%
 \@ifx{#1\undefined}
}%
\providecommand \@ifnum [1]{%
 \ifnum #1\expandafter \@firstoftwo
 \else \expandafter \@secondoftwo
 \fi
}%
\providecommand \@ifx [1]{%
 \ifx #1\expandafter \@firstoftwo
 \else \expandafter \@secondoftwo
 \fi
}%
\providecommand \natexlab [1]{#1}%
\providecommand \enquote  [1]{``#1''}%
\providecommand \bibnamefont  [1]{#1}%
\providecommand \bibfnamefont [1]{#1}%
\providecommand \citenamefont [1]{#1}%
\providecommand \href@noop [0]{\@secondoftwo}%
\providecommand \href [0]{\begingroup \@sanitize@url \@href}%
\providecommand \@href[1]{\@@startlink{#1}\@@href}%
\providecommand \@@href[1]{\endgroup#1\@@endlink}%
\providecommand \@sanitize@url [0]{\catcode `\\12\catcode `\$12\catcode
  `\&12\catcode `\#12\catcode `\^12\catcode `\_12\catcode `\%12\relax}%
\providecommand \@@startlink[1]{}%
\providecommand \@@endlink[0]{}%
\providecommand \url  [0]{\begingroup\@sanitize@url \@url }%
\providecommand \@url [1]{\endgroup\@href {#1}{\urlprefix }}%
\providecommand \urlprefix  [0]{URL }%
\providecommand \Eprint [0]{\href }%
\providecommand \doibase [0]{https://doi.org/}%
\providecommand \selectlanguage [0]{\@gobble}%
\providecommand \bibinfo  [0]{\@secondoftwo}%
\providecommand \bibfield  [0]{\@secondoftwo}%
\providecommand \translation [1]{[#1]}%
\providecommand \BibitemOpen [0]{}%
\providecommand \bibitemStop [0]{}%
\providecommand \bibitemNoStop [0]{.\EOS\space}%
\providecommand \EOS [0]{\spacefactor3000\relax}%
\providecommand \BibitemShut  [1]{\csname bibitem#1\endcsname}%
\let\auto@bib@innerbib\@empty
\bibitem [{\citenamefont {Sakhaee-Pour}\ \emph {et~al.}(2008)\citenamefont
  {Sakhaee-Pour}, \citenamefont {Ahmadian},\ and\ \citenamefont
  {Vafai}}]{Sakhaee-Pour2008-RN84}%
  \BibitemOpen
  \bibfield  {author} {\bibinfo {author} {\bibfnamefont {A.}~\bibnamefont
  {Sakhaee-Pour}}, \bibinfo {author} {\bibfnamefont {M.~T.}\ \bibnamefont
  {Ahmadian}},\ and\ \bibinfo {author} {\bibfnamefont {A.}~\bibnamefont
  {Vafai}},\ }\bibfield  {title} {\bibinfo {title} {Applications of
  single-layered graphene sheets as mass sensors and atomistic dust
  detectors},\ }\href
  {https://doi.org/https://doi.org/10.1016/j.ssc.2007.10.032} {\bibfield
  {journal} {\bibinfo  {journal} {Solid State Communications}\ }\textbf
  {\bibinfo {volume} {145}},\ \bibinfo {pages} {168} (\bibinfo {year}
  {2008})}\BibitemShut {NoStop}%
\bibitem [{\citenamefont {Lassagne}\ \emph {et~al.}(2008)\citenamefont
  {Lassagne}, \citenamefont {Garcia-Sanchez}, \citenamefont {Aguasca},\ and\
  \citenamefont {Bachtold}}]{Lassagne2008-RN44}%
  \BibitemOpen
  \bibfield  {author} {\bibinfo {author} {\bibfnamefont {B.}~\bibnamefont
  {Lassagne}}, \bibinfo {author} {\bibfnamefont {D.}~\bibnamefont
  {Garcia-Sanchez}}, \bibinfo {author} {\bibfnamefont {A.}~\bibnamefont
  {Aguasca}},\ and\ \bibinfo {author} {\bibfnamefont {A.}~\bibnamefont
  {Bachtold}},\ }\bibfield  {title} {\bibinfo {title} {Ultrasensitive mass
  sensing with a nanotube electromechanical resonator},\ }\href
  {https://doi.org/10.1021/nl801982v} {\bibfield  {journal} {\bibinfo
  {journal} {Nano Letters}\ }\textbf {\bibinfo {volume} {8}},\ \bibinfo {pages}
  {3735} (\bibinfo {year} {2008})}\BibitemShut {NoStop}%
\bibitem [{\citenamefont {Atalaya}\ \emph {et~al.}(2010)\citenamefont
  {Atalaya}, \citenamefont {Kinaret},\ and\ \citenamefont
  {Isacsson}}]{Atalaya2010-RN94}%
  \BibitemOpen
  \bibfield  {author} {\bibinfo {author} {\bibfnamefont {J.}~\bibnamefont
  {Atalaya}}, \bibinfo {author} {\bibfnamefont {J.~M.}\ \bibnamefont
  {Kinaret}},\ and\ \bibinfo {author} {\bibfnamefont {A.}~\bibnamefont
  {Isacsson}},\ }\bibfield  {title} {\bibinfo {title} {Nanomechanical mass
  measurement using nonlinear response of a graphene membrane},\ }\href
  {https://doi.org/10.1209/0295-5075/91/48001} {\bibfield  {journal} {\bibinfo
  {journal} {Europhysics Letters}\ }\textbf {\bibinfo {volume} {91}},\ \bibinfo
  {pages} {48001} (\bibinfo {year} {2010})}\BibitemShut {NoStop}%
\bibitem [{\citenamefont {Singh}\ \emph {et~al.}(2010)\citenamefont {Singh},
  \citenamefont {Sengupta}, \citenamefont {Solanki}, \citenamefont {Dhall},
  \citenamefont {Allain}, \citenamefont {Dhara}, \citenamefont {Pant},\ and\
  \citenamefont {Deshmukh}}]{Singh2010-RN4}%
  \BibitemOpen
  \bibfield  {author} {\bibinfo {author} {\bibfnamefont {V.}~\bibnamefont
  {Singh}}, \bibinfo {author} {\bibfnamefont {S.}~\bibnamefont {Sengupta}},
  \bibinfo {author} {\bibfnamefont {H.~S.}\ \bibnamefont {Solanki}}, \bibinfo
  {author} {\bibfnamefont {R.}~\bibnamefont {Dhall}}, \bibinfo {author}
  {\bibfnamefont {A.}~\bibnamefont {Allain}}, \bibinfo {author} {\bibfnamefont
  {S.}~\bibnamefont {Dhara}}, \bibinfo {author} {\bibfnamefont
  {P.}~\bibnamefont {Pant}},\ and\ \bibinfo {author} {\bibfnamefont {M.~M.}\
  \bibnamefont {Deshmukh}},\ }\bibfield  {title} {\bibinfo {title} {Probing
  thermal expansion of graphene and modal dispersion at low-temperature using
  graphene nanoelectromechanical systems resonators},\ }\href
  {https://doi.org/10.1088/0957-4484/21/16/165204} {\bibfield  {journal}
  {\bibinfo  {journal} {Nanotechnology}\ }\textbf {\bibinfo {volume} {21}},\
  \bibinfo {pages} {165204} (\bibinfo {year} {2010})}\BibitemShut {NoStop}%
\bibitem [{\citenamefont {Weber}\ \emph {et~al.}(2016)\citenamefont {Weber},
  \citenamefont {Guttinger}, \citenamefont {Noury}, \citenamefont
  {Vergara-Cruz},\ and\ \citenamefont {Bachtold}}]{Weber2016-RN20}%
  \BibitemOpen
  \bibfield  {author} {\bibinfo {author} {\bibfnamefont {P.}~\bibnamefont
  {Weber}}, \bibinfo {author} {\bibfnamefont {J.}~\bibnamefont {Guttinger}},
  \bibinfo {author} {\bibfnamefont {A.}~\bibnamefont {Noury}}, \bibinfo
  {author} {\bibfnamefont {J.}~\bibnamefont {Vergara-Cruz}},\ and\ \bibinfo
  {author} {\bibfnamefont {A.}~\bibnamefont {Bachtold}},\ }\bibfield  {title}
  {\bibinfo {title} {Force sensitivity of multilayer graphene optomechanical
  devices},\ }\href {https://doi.org/10.1038/ncomms12496} {\bibfield  {journal}
  {\bibinfo  {journal} {Nat Commun}\ }\textbf {\bibinfo {volume} {7}},\
  \bibinfo {pages} {12496} (\bibinfo {year} {2016})}\BibitemShut {NoStop}%
\bibitem [{\citenamefont {Lemme}\ \emph {et~al.}(2020)\citenamefont {Lemme},
  \citenamefont {Wagner}, \citenamefont {Lee}, \citenamefont {Fan},
  \citenamefont {Verbiest}, \citenamefont {Wittmann}, \citenamefont {Lukas},
  \citenamefont {Dolleman}, \citenamefont {Niklaus}, \citenamefont {van~der
  Zant}, \citenamefont {Duesberg},\ and\ \citenamefont
  {Steeneken}}]{Lemme2020-RN3}%
  \BibitemOpen
  \bibfield  {author} {\bibinfo {author} {\bibfnamefont {M.~C.}\ \bibnamefont
  {Lemme}}, \bibinfo {author} {\bibfnamefont {S.}~\bibnamefont {Wagner}},
  \bibinfo {author} {\bibfnamefont {K.}~\bibnamefont {Lee}}, \bibinfo {author}
  {\bibfnamefont {X.}~\bibnamefont {Fan}}, \bibinfo {author} {\bibfnamefont
  {G.~J.}\ \bibnamefont {Verbiest}}, \bibinfo {author} {\bibfnamefont
  {S.}~\bibnamefont {Wittmann}}, \bibinfo {author} {\bibfnamefont
  {S.}~\bibnamefont {Lukas}}, \bibinfo {author} {\bibfnamefont {R.~J.}\
  \bibnamefont {Dolleman}}, \bibinfo {author} {\bibfnamefont {F.}~\bibnamefont
  {Niklaus}}, \bibinfo {author} {\bibfnamefont {H.~S.~J.}\ \bibnamefont
  {van~der Zant}}, \bibinfo {author} {\bibfnamefont {G.~S.}\ \bibnamefont
  {Duesberg}},\ and\ \bibinfo {author} {\bibfnamefont {P.~G.}\ \bibnamefont
  {Steeneken}},\ }\bibfield  {title} {\bibinfo {title} {Nanoelectromechanical
  sensors based on suspended 2d materials},\ }\href
  {https://doi.org/10.34133/2020/8748602} {\bibfield  {journal} {\bibinfo
  {journal} {Research (Wash D C)}\ }\textbf {\bibinfo {volume} {2020}},\
  \bibinfo {pages} {8748602} (\bibinfo {year} {2020})}\BibitemShut {NoStop}%
\bibitem [{\citenamefont {Rosłoń}\ \emph {et~al.}(2022)\citenamefont
  {Rosłoń}, \citenamefont {Japaridze}, \citenamefont {Steeneken},
  \citenamefont {Dekker},\ and\ \citenamefont {Alijani}}]{Roslon2022-RN157}%
  \BibitemOpen
  \bibfield  {author} {\bibinfo {author} {\bibfnamefont {I.~E.}\ \bibnamefont
  {Rosłoń}}, \bibinfo {author} {\bibfnamefont {A.}~\bibnamefont {Japaridze}},
  \bibinfo {author} {\bibfnamefont {P.~G.}\ \bibnamefont {Steeneken}}, \bibinfo
  {author} {\bibfnamefont {C.}~\bibnamefont {Dekker}},\ and\ \bibinfo {author}
  {\bibfnamefont {F.}~\bibnamefont {Alijani}},\ }\bibfield  {title} {\bibinfo
  {title} {Probing nanomotion of single bacteria with graphene drums},\ }\href
  {https://doi.org/10.1038/s41565-022-01111-6} {\bibfield  {journal} {\bibinfo
  {journal} {Nature Nanotechnology}\ }\textbf {\bibinfo {volume} {17}},\
  \bibinfo {pages} {637} (\bibinfo {year} {2022})}\BibitemShut {NoStop}%
\bibitem [{\citenamefont {Romijn}\ \emph {et~al.}(2018)\citenamefont {Romijn},
  \citenamefont {Vollebregt}, \citenamefont {Dolleman}, \citenamefont {Singh},
  \citenamefont {Van Der~Zant}, \citenamefont {Steeneken},\ and\ \citenamefont
  {Sarro}}]{RomijnYear-RN7}%
  \BibitemOpen
  \bibfield  {author} {\bibinfo {author} {\bibfnamefont {J.}~\bibnamefont
  {Romijn}}, \bibinfo {author} {\bibfnamefont {S.}~\bibnamefont {Vollebregt}},
  \bibinfo {author} {\bibfnamefont {R.~J.}\ \bibnamefont {Dolleman}}, \bibinfo
  {author} {\bibfnamefont {M.}~\bibnamefont {Singh}}, \bibinfo {author}
  {\bibfnamefont {H.~S.}\ \bibnamefont {Van Der~Zant}}, \bibinfo {author}
  {\bibfnamefont {P.~G.}\ \bibnamefont {Steeneken}},\ and\ \bibinfo {author}
  {\bibfnamefont {P.~M.}\ \bibnamefont {Sarro}},\ }\bibfield  {title} {\bibinfo
  {title} {A miniaturized low power pirani pressure sensor based on suspended
  graphene},\ }in\ \href {https://doi.org/10.1109/nems.2018.8556902} {\emph
  {\bibinfo {booktitle} {2018 IEEE 13th Annual International Conference on
  Nano/Micro Engineered and Molecular Systems (NEMS)}}}\ (\bibinfo  {publisher}
  {IEEE},\ \bibinfo {year} {2018})\BibitemShut {NoStop}%
\bibitem [{\citenamefont {Dolleman}\ \emph {et~al.}(2016)\citenamefont
  {Dolleman}, \citenamefont {Davidovikj}, \citenamefont {Cartamil-Bueno},
  \citenamefont {Van Der~Zant},\ and\ \citenamefont
  {Steeneken}}]{Dolleman2016-RN88}%
  \BibitemOpen
  \bibfield  {author} {\bibinfo {author} {\bibfnamefont {R.~J.}\ \bibnamefont
  {Dolleman}}, \bibinfo {author} {\bibfnamefont {D.}~\bibnamefont
  {Davidovikj}}, \bibinfo {author} {\bibfnamefont {S.~J.}\ \bibnamefont
  {Cartamil-Bueno}}, \bibinfo {author} {\bibfnamefont {H.~S.~J.}\ \bibnamefont
  {Van Der~Zant}},\ and\ \bibinfo {author} {\bibfnamefont {P.~G.}\ \bibnamefont
  {Steeneken}},\ }\bibfield  {title} {\bibinfo {title} {Graphene squeeze-film
  pressure sensors},\ }\href {https://doi.org/10.1021/acs.nanolett.5b04251}
  {\bibfield  {journal} {\bibinfo  {journal} {Nano Letters}\ }\textbf {\bibinfo
  {volume} {16}},\ \bibinfo {pages} {568} (\bibinfo {year} {2016})}\BibitemShut
  {NoStop}%
\bibitem [{\citenamefont {Shahdeo}\ \emph {et~al.}(2020)\citenamefont
  {Shahdeo}, \citenamefont {Roberts}, \citenamefont {Abbineni},\ and\
  \citenamefont {Gandhi}}]{Shahdeo2020-RN5}%
  \BibitemOpen
  \bibfield  {author} {\bibinfo {author} {\bibfnamefont {D.}~\bibnamefont
  {Shahdeo}}, \bibinfo {author} {\bibfnamefont {A.}~\bibnamefont {Roberts}},
  \bibinfo {author} {\bibfnamefont {N.}~\bibnamefont {Abbineni}},\ and\
  \bibinfo {author} {\bibfnamefont {S.}~\bibnamefont {Gandhi}},\ }\bibinfo
  {title} {Graphene based sensors},\ in\ \href
  {https://doi.org/10.1016/bs.coac.2020.08.007} {\emph {\bibinfo {booktitle}
  {Analytical Applications of Graphene for Comprehensive Analytical
  Chemistry}}},\ \bibinfo {series and number} {Comprehensive Analytical
  Chemistry}\ (\bibinfo {year} {2020})\ pp.\ \bibinfo {pages}
  {175--199}\BibitemShut {NoStop}%
\bibitem [{\citenamefont {Choi}\ \emph {et~al.}(2020)\citenamefont {Choi},
  \citenamefont {Lee}, \citenamefont {Byeon}, \citenamefont {Hong},
  \citenamefont {Park},\ and\ \citenamefont {Lee}}]{Choi2020-RN16}%
  \BibitemOpen
  \bibfield  {author} {\bibinfo {author} {\bibfnamefont {J.~H.}\ \bibnamefont
  {Choi}}, \bibinfo {author} {\bibfnamefont {J.}~\bibnamefont {Lee}}, \bibinfo
  {author} {\bibfnamefont {M.}~\bibnamefont {Byeon}}, \bibinfo {author}
  {\bibfnamefont {T.~E.}\ \bibnamefont {Hong}}, \bibinfo {author}
  {\bibfnamefont {H.}~\bibnamefont {Park}},\ and\ \bibinfo {author}
  {\bibfnamefont {C.~Y.}\ \bibnamefont {Lee}},\ }\bibfield  {title} {\bibinfo
  {title} {Graphene-based gas sensors with high sensitivity and minimal
  sensor-to-sensor variation},\ }\href {https://doi.org/10.1021/acsanm.9b02378}
  {\bibfield  {journal} {\bibinfo  {journal} {ACS Applied Nano Materials}\
  }\textbf {\bibinfo {volume} {3}},\ \bibinfo {pages} {2257} (\bibinfo {year}
  {2020})}\BibitemShut {NoStop}%
\bibitem [{\citenamefont {Dolleman}\ \emph
  {et~al.}(2020{\natexlab{a}})\citenamefont {Dolleman}, \citenamefont
  {Verbiest}, \citenamefont {Blanter}, \citenamefont {Van Der~Zant},\ and\
  \citenamefont {Steeneken}}]{Dolleman2020-RN99}%
  \BibitemOpen
  \bibfield  {author} {\bibinfo {author} {\bibfnamefont {R.~J.}\ \bibnamefont
  {Dolleman}}, \bibinfo {author} {\bibfnamefont {G.~J.}\ \bibnamefont
  {Verbiest}}, \bibinfo {author} {\bibfnamefont {Y.~M.}\ \bibnamefont
  {Blanter}}, \bibinfo {author} {\bibfnamefont {H.~S.~J.}\ \bibnamefont {Van
  Der~Zant}},\ and\ \bibinfo {author} {\bibfnamefont {P.~G.}\ \bibnamefont
  {Steeneken}},\ }\bibfield  {title} {\bibinfo {title} {Nonequilibrium
  thermodynamics of acoustic phonons in suspended graphene},\ }\bibfield
  {journal} {\bibinfo  {journal} {Physical Review Research}\ }\textbf {\bibinfo
  {volume} {2}},\ \href {https://doi.org/10.1103/physrevresearch.2.012058}
  {10.1103/physrevresearch.2.012058} (\bibinfo {year}
  {2020}{\natexlab{a}})\BibitemShut {NoStop}%
\bibitem [{\citenamefont {Liu}\ \emph {et~al.}(2023)\citenamefont {Liu},
  \citenamefont {Baglioni}, \citenamefont {Constant}, \citenamefont {van~der
  Zant}, \citenamefont {Steeneken},\ and\ \citenamefont
  {Verbiest}}]{Hanqing2023-RN180}%
  \BibitemOpen
  \bibfield  {author} {\bibinfo {author} {\bibfnamefont {H.}~\bibnamefont
  {Liu}}, \bibinfo {author} {\bibfnamefont {G.}~\bibnamefont {Baglioni}},
  \bibinfo {author} {\bibfnamefont {C.~B.}\ \bibnamefont {Constant}}, \bibinfo
  {author} {\bibfnamefont {H.~S.}\ \bibnamefont {van~der Zant}}, \bibinfo
  {author} {\bibfnamefont {P.~G.}\ \bibnamefont {Steeneken}},\ and\ \bibinfo
  {author} {\bibfnamefont {G.~J.}\ \bibnamefont {Verbiest}},\ }\bibfield
  {title} {\bibinfo {title} {Enhanced photothermal response near the buckling
  bifurcation in 2d nanomechanical resonators},\ }\href@noop {} {\bibfield
  {journal} {\bibinfo  {journal} {arXiv preprint arXiv:2305.00712}\ } (\bibinfo
  {year} {2023})}\BibitemShut {NoStop}%
\bibitem [{\citenamefont {Sementilli}\ \emph {et~al.}(2022)\citenamefont
  {Sementilli}, \citenamefont {Romero},\ and\ \citenamefont
  {Bowen}}]{Sementilli2022-RN191}%
  \BibitemOpen
  \bibfield  {author} {\bibinfo {author} {\bibfnamefont {L.}~\bibnamefont
  {Sementilli}}, \bibinfo {author} {\bibfnamefont {E.}~\bibnamefont {Romero}},\
  and\ \bibinfo {author} {\bibfnamefont {W.~P.}\ \bibnamefont {Bowen}},\
  }\bibfield  {title} {\bibinfo {title} {Nanomechanical dissipation and strain
  engineering},\ }\href {https://doi.org/10.1002/adfm.202105247} {\bibfield
  {journal} {\bibinfo  {journal} {Advanced Functional Materials}\ }\textbf
  {\bibinfo {volume} {32}},\ \bibinfo {pages} {2105247} (\bibinfo {year}
  {2022})}\BibitemShut {NoStop}%
\bibitem [{\citenamefont {Miller}\ \emph {et~al.}(2018)\citenamefont {Miller},
  \citenamefont {Ansari}, \citenamefont {Heinz}, \citenamefont {Chen},
  \citenamefont {Flader}, \citenamefont {Shin}, \citenamefont {Villanueva},\
  and\ \citenamefont {Kenny}}]{Miller2018-RN128}%
  \BibitemOpen
  \bibfield  {author} {\bibinfo {author} {\bibfnamefont {J.~M.~L.}\
  \bibnamefont {Miller}}, \bibinfo {author} {\bibfnamefont {A.}~\bibnamefont
  {Ansari}}, \bibinfo {author} {\bibfnamefont {D.~B.}\ \bibnamefont {Heinz}},
  \bibinfo {author} {\bibfnamefont {Y.}~\bibnamefont {Chen}}, \bibinfo {author}
  {\bibfnamefont {I.~B.}\ \bibnamefont {Flader}}, \bibinfo {author}
  {\bibfnamefont {D.~D.}\ \bibnamefont {Shin}}, \bibinfo {author}
  {\bibfnamefont {L.~G.}\ \bibnamefont {Villanueva}},\ and\ \bibinfo {author}
  {\bibfnamefont {T.~W.}\ \bibnamefont {Kenny}},\ }\bibfield  {title} {\bibinfo
  {title} {Effective quality factor tuning mechanisms in micromechanical
  resonators},\ }\href {https://doi.org/10.1063/1.5027850} {\bibfield
  {journal} {\bibinfo  {journal} {Applied Physics Reviews}\ }\textbf {\bibinfo
  {volume} {5}},\ \bibinfo {pages} {041307} (\bibinfo {year}
  {2018})}\BibitemShut {NoStop}%
\bibitem [{\citenamefont {Leeson}(1966)}]{Leeson1966-RN213}%
  \BibitemOpen
  \bibfield  {author} {\bibinfo {author} {\bibfnamefont {D.}~\bibnamefont
  {Leeson}},\ }\bibfield  {title} {\bibinfo {title} {A simple model of feedback
  oscillator noise spectrum},\ }\href {https://doi.org/10.1109/proc.1966.4682}
  {\bibfield  {journal} {\bibinfo  {journal} {Proceedings of the IEEE}\
  }\textbf {\bibinfo {volume} {54}},\ \bibinfo {pages} {329} (\bibinfo {year}
  {1966})}\BibitemShut {NoStop}%
\bibitem [{\citenamefont {Gabrielson}(1995)}]{Gabrielson1995-RN210}%
  \BibitemOpen
  \bibfield  {author} {\bibinfo {author} {\bibfnamefont {T.~B.}\ \bibnamefont
  {Gabrielson}},\ }\bibfield  {title} {\bibinfo {title} {Fundamental noise
  limits for miniature acoustic and vibration sensors},\ }\href
  {https://doi.org/10.1115/1.2874471} {\bibfield  {journal} {\bibinfo
  {journal} {Journal of Vibration and Acoustics}\ }\textbf {\bibinfo {volume}
  {117}},\ \bibinfo {pages} {405} (\bibinfo {year} {1995})}\BibitemShut
  {NoStop}%
\bibitem [{\citenamefont {Gabrielson}(1993)}]{Gabrielson1993-RN211}%
  \BibitemOpen
  \bibfield  {author} {\bibinfo {author} {\bibfnamefont {T.}~\bibnamefont
  {Gabrielson}},\ }\bibfield  {title} {\bibinfo {title} {Mechanical-thermal
  noise in micromachined acoustic and vibration sensors},\ }\href
  {https://doi.org/10.1109/16.210197} {\bibfield  {journal} {\bibinfo
  {journal} {IEEE Transactions on Electron Devices}\ }\textbf {\bibinfo
  {volume} {40}},\ \bibinfo {pages} {903} (\bibinfo {year} {1993})}\BibitemShut
  {NoStop}%
\bibitem [{\citenamefont {Schmid}(2016)}]{Schmid2016-RN203}%
  \BibitemOpen
  \bibfield  {author} {\bibinfo {author} {\bibfnamefont {S.}~\bibnamefont
  {Schmid}},\ }\href@noop {} {\emph {\bibinfo {title} {Fundamentals of
  Nanomechanical Resonators}}}\ (\bibinfo {year} {2016})\BibitemShut {NoStop}%
\bibitem [{\citenamefont {Verbridge}\ \emph {et~al.}(2006)\citenamefont
  {Verbridge}, \citenamefont {Parpia}, \citenamefont {Reichenbach},
  \citenamefont {Bellan},\ and\ \citenamefont
  {Craighead}}]{Verbridge2006-RN200}%
  \BibitemOpen
  \bibfield  {author} {\bibinfo {author} {\bibfnamefont {S.~S.}\ \bibnamefont
  {Verbridge}}, \bibinfo {author} {\bibfnamefont {J.~M.}\ \bibnamefont
  {Parpia}}, \bibinfo {author} {\bibfnamefont {R.~B.}\ \bibnamefont
  {Reichenbach}}, \bibinfo {author} {\bibfnamefont {L.~M.}\ \bibnamefont
  {Bellan}},\ and\ \bibinfo {author} {\bibfnamefont {H.~G.}\ \bibnamefont
  {Craighead}},\ }\bibfield  {title} {\bibinfo {title} {High quality factor
  resonance at room temperature with nanostrings under high tensile stress},\
  }\href@noop {} {\bibfield  {journal} {\bibinfo  {journal} {Journal of Applied
  Physics}\ }\textbf {\bibinfo {volume} {99}},\ \bibinfo {pages} {124304}
  (\bibinfo {year} {2006})}\BibitemShut {NoStop}%
\bibitem [{\citenamefont {Fedorov}\ \emph {et~al.}(2019)\citenamefont
  {Fedorov}, \citenamefont {Engelsen}, \citenamefont {Ghadimi}, \citenamefont
  {Bereyhi}, \citenamefont {Schilling}, \citenamefont {Wilson},\ and\
  \citenamefont {Kippenberg}}]{Fedorov2019-RN218}%
  \BibitemOpen
  \bibfield  {author} {\bibinfo {author} {\bibfnamefont {S.~A.}\ \bibnamefont
  {Fedorov}}, \bibinfo {author} {\bibfnamefont {N.~J.}\ \bibnamefont
  {Engelsen}}, \bibinfo {author} {\bibfnamefont {A.~H.}\ \bibnamefont
  {Ghadimi}}, \bibinfo {author} {\bibfnamefont {M.~J.}\ \bibnamefont
  {Bereyhi}}, \bibinfo {author} {\bibfnamefont {R.}~\bibnamefont {Schilling}},
  \bibinfo {author} {\bibfnamefont {D.~J.}\ \bibnamefont {Wilson}},\ and\
  \bibinfo {author} {\bibfnamefont {T.~J.}\ \bibnamefont {Kippenberg}},\
  }\bibfield  {title} {\bibinfo {title} {Generalized dissipation dilution in
  strained mechanical resonators},\ }\bibfield  {journal} {\bibinfo  {journal}
  {Physical Review B}\ }\textbf {\bibinfo {volume} {99}},\ \href
  {https://doi.org/10.1103/physrevb.99.054107} {10.1103/physrevb.99.054107}
  (\bibinfo {year} {2019})\BibitemShut {NoStop}%
\bibitem [{\citenamefont {Davidovikj}\ \emph {et~al.}(2018)\citenamefont
  {Davidovikj}, \citenamefont {Poot}, \citenamefont {Cartamil-Bueno},
  \citenamefont {Van Der~Zant},\ and\ \citenamefont
  {Steeneken}}]{Davidovikj2018-RN96}%
  \BibitemOpen
  \bibfield  {author} {\bibinfo {author} {\bibfnamefont {D.}~\bibnamefont
  {Davidovikj}}, \bibinfo {author} {\bibfnamefont {M.}~\bibnamefont {Poot}},
  \bibinfo {author} {\bibfnamefont {S.~J.}\ \bibnamefont {Cartamil-Bueno}},
  \bibinfo {author} {\bibfnamefont {H.~S.~J.}\ \bibnamefont {Van Der~Zant}},\
  and\ \bibinfo {author} {\bibfnamefont {P.~G.}\ \bibnamefont {Steeneken}},\
  }\bibfield  {title} {\bibinfo {title} {On-chip heaters for tension tuning of
  graphene nanodrums},\ }\href {https://doi.org/10.1021/acs.nanolett.7b05358}
  {\bibfield  {journal} {\bibinfo  {journal} {Nano Letters}\ }\textbf {\bibinfo
  {volume} {18}},\ \bibinfo {pages} {2852} (\bibinfo {year}
  {2018})}\BibitemShut {NoStop}%
\bibitem [{\citenamefont {Morell}\ \emph {et~al.}(2016)\citenamefont {Morell},
  \citenamefont {Reserbat-Plantey}, \citenamefont {Tsioutsios}, \citenamefont
  {Schädler}, \citenamefont {Dubin}, \citenamefont {Koppens},\ and\
  \citenamefont {Bachtold}}]{Morell2016-RN228}%
  \BibitemOpen
  \bibfield  {author} {\bibinfo {author} {\bibfnamefont {N.}~\bibnamefont
  {Morell}}, \bibinfo {author} {\bibfnamefont {A.}~\bibnamefont
  {Reserbat-Plantey}}, \bibinfo {author} {\bibfnamefont {I.}~\bibnamefont
  {Tsioutsios}}, \bibinfo {author} {\bibfnamefont {K.~G.}\ \bibnamefont
  {Schädler}}, \bibinfo {author} {\bibfnamefont {F.}~\bibnamefont {Dubin}},
  \bibinfo {author} {\bibfnamefont {F.~H.~L.}\ \bibnamefont {Koppens}},\ and\
  \bibinfo {author} {\bibfnamefont {A.}~\bibnamefont {Bachtold}},\ }\bibfield
  {title} {\bibinfo {title} {High quality factor mechanical resonators based on
  wse<sub>2</sub> monolayers},\ }\href
  {https://doi.org/10.1021/acs.nanolett.6b02038} {\bibfield  {journal}
  {\bibinfo  {journal} {Nano Letters}\ }\textbf {\bibinfo {volume} {16}},\
  \bibinfo {pages} {5102} (\bibinfo {year} {2016})}\BibitemShut {NoStop}%
\bibitem [{\citenamefont {Song}\ \emph {et~al.}(2012)\citenamefont {Song},
  \citenamefont {Oksanen}, \citenamefont {Sillanpää}, \citenamefont
  {Craighead}, \citenamefont {Parpia},\ and\ \citenamefont
  {Hakonen}}]{Song2012-RN140}%
  \BibitemOpen
  \bibfield  {author} {\bibinfo {author} {\bibfnamefont {X.}~\bibnamefont
  {Song}}, \bibinfo {author} {\bibfnamefont {M.}~\bibnamefont {Oksanen}},
  \bibinfo {author} {\bibfnamefont {M.~A.}\ \bibnamefont {Sillanpää}},
  \bibinfo {author} {\bibfnamefont {H.~G.}\ \bibnamefont {Craighead}}, \bibinfo
  {author} {\bibfnamefont {J.~M.}\ \bibnamefont {Parpia}},\ and\ \bibinfo
  {author} {\bibfnamefont {P.~J.}\ \bibnamefont {Hakonen}},\ }\bibfield
  {title} {\bibinfo {title} {Stamp transferred suspended graphene mechanical
  resonators for radio frequency electrical readout},\ }\href
  {https://doi.org/10.1021/nl203305q} {\bibfield  {journal} {\bibinfo
  {journal} {Nano Letters}\ }\textbf {\bibinfo {volume} {12}},\ \bibinfo
  {pages} {198} (\bibinfo {year} {2012})}\BibitemShut {NoStop}%
\bibitem [{\citenamefont {Rieger}\ \emph {et~al.}(2014)\citenamefont {Rieger},
  \citenamefont {Isacsson}, \citenamefont {Seitner}, \citenamefont {Kotthaus},\
  and\ \citenamefont {Weig}}]{Rieger2014-RN229}%
  \BibitemOpen
  \bibfield  {author} {\bibinfo {author} {\bibfnamefont {J.}~\bibnamefont
  {Rieger}}, \bibinfo {author} {\bibfnamefont {A.}~\bibnamefont {Isacsson}},
  \bibinfo {author} {\bibfnamefont {M.~J.}\ \bibnamefont {Seitner}}, \bibinfo
  {author} {\bibfnamefont {J.~P.}\ \bibnamefont {Kotthaus}},\ and\ \bibinfo
  {author} {\bibfnamefont {E.~M.}\ \bibnamefont {Weig}},\ }\bibfield  {title}
  {\bibinfo {title} {Energy losses of nanomechanical resonators induced by
  atomic force microscopy-controlled mechanical impedance mismatching},\
  }\bibfield  {journal} {\bibinfo  {journal} {Nature Communications}\ }\textbf
  {\bibinfo {volume} {5}},\ \href {https://doi.org/10.1038/ncomms4345}
  {10.1038/ncomms4345} (\bibinfo {year} {2014})\BibitemShut {NoStop}%
\bibitem [{\citenamefont {Steeneken}\ \emph {et~al.}(2021)\citenamefont
  {Steeneken}, \citenamefont {Dolleman}, \citenamefont {Davidovikj},
  \citenamefont {Alijani},\ and\ \citenamefont {van~der
  Zant}}]{Steeneken2021-RN1}%
  \BibitemOpen
  \bibfield  {author} {\bibinfo {author} {\bibfnamefont {P.~G.}\ \bibnamefont
  {Steeneken}}, \bibinfo {author} {\bibfnamefont {R.~J.}\ \bibnamefont
  {Dolleman}}, \bibinfo {author} {\bibfnamefont {D.}~\bibnamefont
  {Davidovikj}}, \bibinfo {author} {\bibfnamefont {F.}~\bibnamefont
  {Alijani}},\ and\ \bibinfo {author} {\bibfnamefont {H.~S.~J.}\ \bibnamefont
  {van~der Zant}},\ }\bibfield  {title} {\bibinfo {title} {Dynamics of 2d
  material membranes},\ }\bibfield  {journal} {\bibinfo  {journal} {2D
  Materials}\ }\textbf {\bibinfo {volume} {8}},\ \href
  {https://doi.org/10.1088/2053-1583/ac152c} {10.1088/2053-1583/ac152c}
  (\bibinfo {year} {2021})\BibitemShut {NoStop}%
\bibitem [{\citenamefont {Lifshitz}\ and\ \citenamefont
  {Roukes}(2000)}]{Liftshitz2000-RN193}%
  \BibitemOpen
  \bibfield  {author} {\bibinfo {author} {\bibfnamefont {R.}~\bibnamefont
  {Lifshitz}}\ and\ \bibinfo {author} {\bibfnamefont {M.~L.}\ \bibnamefont
  {Roukes}},\ }\bibfield  {title} {\bibinfo {title} {Thermoelastic damping in
  micro- and nanomechanical systems},\ }\href
  {https://doi.org/10.1103/physrevb.61.5600} {\bibfield  {journal} {\bibinfo
  {journal} {Physical Review B}\ }\textbf {\bibinfo {volume} {61}},\ \bibinfo
  {pages} {5600} (\bibinfo {year} {2000})}\BibitemShut {NoStop}%
\bibitem [{\citenamefont {Verbiest}\ \emph {et~al.}(2016)\citenamefont
  {Verbiest}, \citenamefont {Xu}, \citenamefont {Goldsche}, \citenamefont
  {Khodkov}, \citenamefont {Barzanjeh}, \citenamefont {Von Den~Driesch},
  \citenamefont {Buca},\ and\ \citenamefont {Stampfer}}]{Verbiest2016-RN11}%
  \BibitemOpen
  \bibfield  {author} {\bibinfo {author} {\bibfnamefont {G.~J.}\ \bibnamefont
  {Verbiest}}, \bibinfo {author} {\bibfnamefont {D.}~\bibnamefont {Xu}},
  \bibinfo {author} {\bibfnamefont {M.}~\bibnamefont {Goldsche}}, \bibinfo
  {author} {\bibfnamefont {T.}~\bibnamefont {Khodkov}}, \bibinfo {author}
  {\bibfnamefont {S.}~\bibnamefont {Barzanjeh}}, \bibinfo {author}
  {\bibfnamefont {N.}~\bibnamefont {Von Den~Driesch}}, \bibinfo {author}
  {\bibfnamefont {D.}~\bibnamefont {Buca}},\ and\ \bibinfo {author}
  {\bibfnamefont {C.}~\bibnamefont {Stampfer}},\ }\bibfield  {title} {\bibinfo
  {title} {Tunable mechanical coupling between driven microelectromechanical
  resonators},\ }\href {https://doi.org/10.1063/1.4964122} {\bibfield
  {journal} {\bibinfo  {journal} {Applied Physics Letters}\ }\textbf {\bibinfo
  {volume} {109}},\ \bibinfo {pages} {143507} (\bibinfo {year}
  {2016})}\BibitemShut {NoStop}%
\bibitem [{\citenamefont {Perez-Garza}\ \emph {et~al.}(2014)\citenamefont
  {Perez-Garza}, \citenamefont {Kievit}, \citenamefont {Schneider},\ and\
  \citenamefont {Staufer}}]{Perez-Garza2014-RN9}%
  \BibitemOpen
  \bibfield  {author} {\bibinfo {author} {\bibfnamefont {H.~H.}\ \bibnamefont
  {Perez-Garza}}, \bibinfo {author} {\bibfnamefont {E.~W.}\ \bibnamefont
  {Kievit}}, \bibinfo {author} {\bibfnamefont {G.~F.}\ \bibnamefont
  {Schneider}},\ and\ \bibinfo {author} {\bibfnamefont {U.}~\bibnamefont
  {Staufer}},\ }\bibfield  {title} {\bibinfo {title} {Highly strained graphene
  samples of varying thickness and comparison of their behaviour},\ }\href
  {https://doi.org/10.1088/0957-4484/25/46/465708} {\bibfield  {journal}
  {\bibinfo  {journal} {Nanotechnology}\ }\textbf {\bibinfo {volume} {25}},\
  \bibinfo {pages} {465708} (\bibinfo {year} {2014})}\BibitemShut {NoStop}%
\bibitem [{\citenamefont {Xie}\ \emph {et~al.}(2021)\citenamefont {Xie},
  \citenamefont {Lee}, \citenamefont {Wang},\ and\ \citenamefont
  {Feng}}]{Xie2021-RN59}%
  \BibitemOpen
  \bibfield  {author} {\bibinfo {author} {\bibfnamefont {Y.}~\bibnamefont
  {Xie}}, \bibinfo {author} {\bibfnamefont {J.}~\bibnamefont {Lee}}, \bibinfo
  {author} {\bibfnamefont {Y.}~\bibnamefont {Wang}},\ and\ \bibinfo {author}
  {\bibfnamefont {P.~X.}\ \bibnamefont {Feng}},\ }\bibfield  {title} {\bibinfo
  {title} {Nanoelectromechanical systems: Straining and tuning atomic layer
  nanoelectromechanical resonators via comb‐drive mems actuators (adv. mater.
  technol. 2/2021)},\ }\href {https://doi.org/10.1002/admt.202170008}
  {\bibfield  {journal} {\bibinfo  {journal} {Advanced Materials Technologies}\
  }\textbf {\bibinfo {volume} {6}},\ \bibinfo {pages} {2000794} (\bibinfo
  {year} {2021})}\BibitemShut {NoStop}%
\bibitem [{\citenamefont {Zou}\ \emph {et~al.}()\citenamefont {Zou},
  \citenamefont {Ahmed}, \citenamefont {Jaber},\ and\ \citenamefont
  {Fariborzi}}]{ZouYear-RN231}%
  \BibitemOpen
  \bibfield  {author} {\bibinfo {author} {\bibfnamefont {X.}~\bibnamefont
  {Zou}}, \bibinfo {author} {\bibfnamefont {S.}~\bibnamefont {Ahmed}}, \bibinfo
  {author} {\bibfnamefont {N.}~\bibnamefont {Jaber}},\ and\ \bibinfo {author}
  {\bibfnamefont {H.}~\bibnamefont {Fariborzi}},\ }\bibfield  {title} {\bibinfo
  {title} {A compact high-sensitivity temperature sensor using an encapsulated
  clamped-clamped mems beam resonator},\ }in\ \href
  {https://doi.org/10.1109/transducers50396.2021.9495651} {\emph {\bibinfo
  {booktitle} {2021 21st International Conference on Solid-State Sensors,
  Actuators and Microsystems (Transducers)}}}\ (\bibinfo  {publisher}
  {IEEE})\BibitemShut {NoStop}%
\bibitem [{\citenamefont {Streit}\ \emph {et~al.}()\citenamefont {Streit},
  \citenamefont {Forke}, \citenamefont {Voigt}, \citenamefont {Schwarz},
  \citenamefont {Ziegenhardt}, \citenamefont {Weidlich}, \citenamefont
  {Billep}, \citenamefont {Gaitzsch},\ and\ \citenamefont
  {Kuhn}}]{StreitYear-RN230}%
  \BibitemOpen
  \bibfield  {author} {\bibinfo {author} {\bibfnamefont {P.}~\bibnamefont
  {Streit}}, \bibinfo {author} {\bibfnamefont {R.}~\bibnamefont {Forke}},
  \bibinfo {author} {\bibfnamefont {S.}~\bibnamefont {Voigt}}, \bibinfo
  {author} {\bibfnamefont {U.}~\bibnamefont {Schwarz}}, \bibinfo {author}
  {\bibfnamefont {R.}~\bibnamefont {Ziegenhardt}}, \bibinfo {author}
  {\bibfnamefont {S.}~\bibnamefont {Weidlich}}, \bibinfo {author}
  {\bibfnamefont {D.}~\bibnamefont {Billep}}, \bibinfo {author} {\bibfnamefont
  {M.}~\bibnamefont {Gaitzsch}},\ and\ \bibinfo {author} {\bibfnamefont
  {H.}~\bibnamefont {Kuhn}},\ }\bibfield  {title} {\bibinfo {title} {Vibration
  sensors with a high bandwidth and low snr, enhanced with post processing gap
  reduction},\ }in\ \href {https://doi.org/10.1109/eurosime54907.2022.9758891}
  {\emph {\bibinfo {booktitle} {2022 23rd International Conference on Thermal,
  Mechanical and Multi-Physics Simulation and Experiments in Microelectronics
  and Microsystems (EuroSimE)}}}\ (\bibinfo  {publisher} {IEEE})\BibitemShut
  {NoStop}%
\bibitem [{\citenamefont {Novoselov}\ \emph {et~al.}(2005)\citenamefont
  {Novoselov}, \citenamefont {Jiang}, \citenamefont {Schedin}, \citenamefont
  {Booth}, \citenamefont {Khotkevich}, \citenamefont {Morozov},\ and\
  \citenamefont {Geim}}]{Novoselov2005-RN222}%
  \BibitemOpen
  \bibfield  {author} {\bibinfo {author} {\bibfnamefont {K.~S.}\ \bibnamefont
  {Novoselov}}, \bibinfo {author} {\bibfnamefont {D.}~\bibnamefont {Jiang}},
  \bibinfo {author} {\bibfnamefont {F.}~\bibnamefont {Schedin}}, \bibinfo
  {author} {\bibfnamefont {T.~J.}\ \bibnamefont {Booth}}, \bibinfo {author}
  {\bibfnamefont {V.~V.}\ \bibnamefont {Khotkevich}}, \bibinfo {author}
  {\bibfnamefont {S.~V.}\ \bibnamefont {Morozov}},\ and\ \bibinfo {author}
  {\bibfnamefont {A.~K.}\ \bibnamefont {Geim}},\ }\bibfield  {title} {\bibinfo
  {title} {Two-dimensional atomic crystals},\ }\href
  {https://doi.org/10.1073/pnas.0502848102} {\bibfield  {journal} {\bibinfo
  {journal} {Proceedings of the National Academy of Sciences}\ }\textbf
  {\bibinfo {volume} {102}},\ \bibinfo {pages} {10451} (\bibinfo {year}
  {2005})}\BibitemShut {NoStop}%
\bibitem [{\citenamefont {Kinoshita}\ \emph {et~al.}(2019)\citenamefont
  {Kinoshita}, \citenamefont {Moriya}, \citenamefont {Onodera}, \citenamefont
  {Wakafuji}, \citenamefont {Masubuchi}, \citenamefont {Watanabe},
  \citenamefont {Taniguchi},\ and\ \citenamefont
  {Machida}}]{Kinoshita2019-RN41}%
  \BibitemOpen
  \bibfield  {author} {\bibinfo {author} {\bibfnamefont {K.}~\bibnamefont
  {Kinoshita}}, \bibinfo {author} {\bibfnamefont {R.}~\bibnamefont {Moriya}},
  \bibinfo {author} {\bibfnamefont {M.}~\bibnamefont {Onodera}}, \bibinfo
  {author} {\bibfnamefont {Y.}~\bibnamefont {Wakafuji}}, \bibinfo {author}
  {\bibfnamefont {S.}~\bibnamefont {Masubuchi}}, \bibinfo {author}
  {\bibfnamefont {K.}~\bibnamefont {Watanabe}}, \bibinfo {author}
  {\bibfnamefont {T.}~\bibnamefont {Taniguchi}},\ and\ \bibinfo {author}
  {\bibfnamefont {T.}~\bibnamefont {Machida}},\ }\bibfield  {title} {\bibinfo
  {title} {Dry release transfer of graphene and few-layer h-bn by utilizing
  thermoplasticity of polypropylene carbonate},\ }\bibfield  {journal}
  {\bibinfo  {journal} {npj 2D Materials and Applications}\ }\textbf {\bibinfo
  {volume} {3}},\ \href {https://doi.org/10.1038/s41699-019-0104-8}
  {10.1038/s41699-019-0104-8} (\bibinfo {year} {2019})\BibitemShut {NoStop}%
\bibitem [{\citenamefont {Wang}\ \emph {et~al.}(1997)\citenamefont {Wang},
  \citenamefont {Huang}, \citenamefont {Liao}, \citenamefont {Lin},
  \citenamefont {Cong},\ and\ \citenamefont {Chen}}]{Wang1997-RN189}%
  \BibitemOpen
  \bibfield  {author} {\bibinfo {author} {\bibfnamefont {S.}~\bibnamefont
  {Wang}}, \bibinfo {author} {\bibfnamefont {Y.}~\bibnamefont {Huang}},
  \bibinfo {author} {\bibfnamefont {B.}~\bibnamefont {Liao}}, \bibinfo {author}
  {\bibfnamefont {G.}~\bibnamefont {Lin}}, \bibinfo {author} {\bibfnamefont
  {G.}~\bibnamefont {Cong}},\ and\ \bibinfo {author} {\bibfnamefont
  {L.}~\bibnamefont {Chen}},\ }\bibfield  {title} {\bibinfo {title} {Structure
  and properties of poly(propylene carbonate)},\ }\href
  {https://doi.org/10.1080/10236669708032759} {\bibfield  {journal} {\bibinfo
  {journal} {International Journal of Polymer Analysis and Characterization}\
  }\textbf {\bibinfo {volume} {3}},\ \bibinfo {pages} {131} (\bibinfo {year}
  {1997})}\BibitemShut {NoStop}%
\bibitem [{\citenamefont {Lee}\ \emph {et~al.}(2019)\citenamefont {Lee},
  \citenamefont {Davidovikj}, \citenamefont {Sajadi}, \citenamefont {Siskins},
  \citenamefont {Alijani}, \citenamefont {van~der Zant},\ and\ \citenamefont
  {Steeneken}}]{Lee2019-RN60}%
  \BibitemOpen
  \bibfield  {author} {\bibinfo {author} {\bibfnamefont {M.}~\bibnamefont
  {Lee}}, \bibinfo {author} {\bibfnamefont {D.}~\bibnamefont {Davidovikj}},
  \bibinfo {author} {\bibfnamefont {B.}~\bibnamefont {Sajadi}}, \bibinfo
  {author} {\bibfnamefont {M.}~\bibnamefont {Siskins}}, \bibinfo {author}
  {\bibfnamefont {F.}~\bibnamefont {Alijani}}, \bibinfo {author} {\bibfnamefont
  {H.~S.~J.}\ \bibnamefont {van~der Zant}},\ and\ \bibinfo {author}
  {\bibfnamefont {P.~G.}\ \bibnamefont {Steeneken}},\ }\bibfield  {title}
  {\bibinfo {title} {Sealing graphene nanodrums},\ }\href
  {https://doi.org/10.1021/acs.nanolett.9b01770} {\bibfield  {journal}
  {\bibinfo  {journal} {Nano Lett}\ }\textbf {\bibinfo {volume} {19}},\
  \bibinfo {pages} {5313} (\bibinfo {year} {2019})}\BibitemShut {NoStop}%
\bibitem [{\citenamefont {Nicholl}\ \emph {et~al.}(2015)\citenamefont
  {Nicholl}, \citenamefont {Conley}, \citenamefont {Lavrik}, \citenamefont
  {Vlassiouk}, \citenamefont {Puzyrev}, \citenamefont {Sreenivas},
  \citenamefont {Pantelides},\ and\ \citenamefont
  {Bolotin}}]{Nicholl2015-RN119}%
  \BibitemOpen
  \bibfield  {author} {\bibinfo {author} {\bibfnamefont {R.~J.}\ \bibnamefont
  {Nicholl}}, \bibinfo {author} {\bibfnamefont {H.~J.}\ \bibnamefont {Conley}},
  \bibinfo {author} {\bibfnamefont {N.~V.}\ \bibnamefont {Lavrik}}, \bibinfo
  {author} {\bibfnamefont {I.}~\bibnamefont {Vlassiouk}}, \bibinfo {author}
  {\bibfnamefont {Y.~S.}\ \bibnamefont {Puzyrev}}, \bibinfo {author}
  {\bibfnamefont {V.~P.}\ \bibnamefont {Sreenivas}}, \bibinfo {author}
  {\bibfnamefont {S.~T.}\ \bibnamefont {Pantelides}},\ and\ \bibinfo {author}
  {\bibfnamefont {K.~I.}\ \bibnamefont {Bolotin}},\ }\bibfield  {title}
  {\bibinfo {title} {The effect of intrinsic crumpling on the mechanics of
  free-standing graphene},\ }\href {https://doi.org/10.1038/ncomms9789}
  {\bibfield  {journal} {\bibinfo  {journal} {Nature Communications}\ }\textbf
  {\bibinfo {volume} {6}},\ \bibinfo {pages} {8789} (\bibinfo {year}
  {2015})}\BibitemShut {NoStop}%
\bibitem [{\citenamefont {Schmidt}\ \emph {et~al.}(2020)\citenamefont
  {Schmidt}, \citenamefont {Schitter},\ and\ \citenamefont
  {Rankers}}]{Schmidt2020-RN196}%
  \BibitemOpen
  \bibfield  {author} {\bibinfo {author} {\bibfnamefont {R.~M.}\ \bibnamefont
  {Schmidt}}, \bibinfo {author} {\bibfnamefont {G.}~\bibnamefont {Schitter}},\
  and\ \bibinfo {author} {\bibfnamefont {A.}~\bibnamefont {Rankers}},\
  }\href@noop {} {\emph {\bibinfo {title} {The design of high performance
  mechatronics-: high-Tech functionality by multidisciplinary system
  integration}}}\ (\bibinfo  {publisher} {Ios Press},\ \bibinfo {year}
  {2020})\BibitemShut {NoStop}%
\bibitem [{\citenamefont {Figliola}\ and\ \citenamefont
  {Beasley}(2014)}]{Figliola2014-RN197}%
  \BibitemOpen
  \bibfield  {author} {\bibinfo {author} {\bibfnamefont {R.~S.}\ \bibnamefont
  {Figliola}}\ and\ \bibinfo {author} {\bibfnamefont {D.~E.}\ \bibnamefont
  {Beasley}},\ }\href@noop {} {\emph {\bibinfo {title} {Theory and design for
  mechanical measurements}}}\ (\bibinfo  {publisher} {John Wiley \& Sons},\
  \bibinfo {year} {2014})\BibitemShut {NoStop}%
\bibitem [{\citenamefont {Verbiest}\ \emph {et~al.}(2021)\citenamefont
  {Verbiest}, \citenamefont {Goldsche}, \citenamefont {Sonntag}, \citenamefont
  {Khodkov}, \citenamefont {von~den Driesch}, \citenamefont {Buca},\ and\
  \citenamefont {Stampfer}}]{Verbiest2021-RN15}%
  \BibitemOpen
  \bibfield  {author} {\bibinfo {author} {\bibfnamefont {G.~J.}\ \bibnamefont
  {Verbiest}}, \bibinfo {author} {\bibfnamefont {M.}~\bibnamefont {Goldsche}},
  \bibinfo {author} {\bibfnamefont {J.}~\bibnamefont {Sonntag}}, \bibinfo
  {author} {\bibfnamefont {T.}~\bibnamefont {Khodkov}}, \bibinfo {author}
  {\bibfnamefont {N.}~\bibnamefont {von~den Driesch}}, \bibinfo {author}
  {\bibfnamefont {D.}~\bibnamefont {Buca}},\ and\ \bibinfo {author}
  {\bibfnamefont {C.}~\bibnamefont {Stampfer}},\ }\bibfield  {title} {\bibinfo
  {title} {Tunable coupling of two mechanical resonators by a graphene
  membrane},\ }\href {https://doi.org/10.1088/2053-1583/ac005e} {\bibfield
  {journal} {\bibinfo  {journal} {2D Materials}\ }\textbf {\bibinfo {volume}
  {8}},\ \bibinfo {pages} {035039} (\bibinfo {year} {2021})}\BibitemShut
  {NoStop}%
\bibitem [{\citenamefont {Goldsche}\ \emph
  {et~al.}(2018{\natexlab{a}})\citenamefont {Goldsche}, \citenamefont
  {Verbiest}, \citenamefont {Khodkov}, \citenamefont {Sonntag}, \citenamefont
  {Driesch}, \citenamefont {Buca},\ and\ \citenamefont
  {Stampfer}}]{Goldsche2018-RN40}%
  \BibitemOpen
  \bibfield  {author} {\bibinfo {author} {\bibfnamefont {M.}~\bibnamefont
  {Goldsche}}, \bibinfo {author} {\bibfnamefont {G.~J.}\ \bibnamefont
  {Verbiest}}, \bibinfo {author} {\bibfnamefont {T.}~\bibnamefont {Khodkov}},
  \bibinfo {author} {\bibfnamefont {J.}~\bibnamefont {Sonntag}}, \bibinfo
  {author} {\bibfnamefont {N.~V.~D.}\ \bibnamefont {Driesch}}, \bibinfo
  {author} {\bibfnamefont {D.}~\bibnamefont {Buca}},\ and\ \bibinfo {author}
  {\bibfnamefont {C.}~\bibnamefont {Stampfer}},\ }\bibfield  {title} {\bibinfo
  {title} {Fabrication of comb-drive actuators for straining nanostructured
  suspended graphene},\ }\href {https://doi.org/10.1088/1361-6528/aacdec}
  {\bibfield  {journal} {\bibinfo  {journal} {Nanotechnology}\ }\textbf
  {\bibinfo {volume} {29}},\ \bibinfo {pages} {375301} (\bibinfo {year}
  {2018}{\natexlab{a}})}\BibitemShut {NoStop}%
\bibitem [{\citenamefont {Sonntag}\ \emph {et~al.}()\citenamefont {Sonntag},
  \citenamefont {Goldsche}, \citenamefont {Khodkov}, \citenamefont {Verbiest},
  \citenamefont {Reichardt}, \citenamefont {Den~Driesch}, \citenamefont
  {Buca},\ and\ \citenamefont {Stampfer}}]{SonntagYear-RN122}%
  \BibitemOpen
  \bibfield  {author} {\bibinfo {author} {\bibfnamefont {J.}~\bibnamefont
  {Sonntag}}, \bibinfo {author} {\bibfnamefont {M.}~\bibnamefont {Goldsche}},
  \bibinfo {author} {\bibfnamefont {T.}~\bibnamefont {Khodkov}}, \bibinfo
  {author} {\bibfnamefont {G.}~\bibnamefont {Verbiest}}, \bibinfo {author}
  {\bibfnamefont {S.}~\bibnamefont {Reichardt}}, \bibinfo {author}
  {\bibfnamefont {N.~V.}\ \bibnamefont {Den~Driesch}}, \bibinfo {author}
  {\bibfnamefont {D.}~\bibnamefont {Buca}},\ and\ \bibinfo {author}
  {\bibfnamefont {C.}~\bibnamefont {Stampfer}},\ }\bibfield  {title} {\bibinfo
  {title} {Engineering tunable strain fields in suspended graphene by
  microelectromechanical systems},\ }in\ \href
  {https://doi.org/10.1109/transducers.2019.8808807} {\emph {\bibinfo
  {booktitle} {2019 20th International Conference on Solid-State Sensors,
  Actuators and Microsystems \& Eurosensors XXXIII (TRANSDUCERS \& EUROSENSORS
  XXXIII)}}}\ (\bibinfo  {publisher} {IEEE})\BibitemShut {NoStop}%
\bibitem [{\citenamefont {Ying}\ \emph {et~al.}(2022)\citenamefont {Ying},
  \citenamefont {Zhang}, \citenamefont {Moser}, \citenamefont {Su},
  \citenamefont {Song},\ and\ \citenamefont {Guo}}]{Ying2022-RN53}%
  \BibitemOpen
  \bibfield  {author} {\bibinfo {author} {\bibfnamefont {Y.}~\bibnamefont
  {Ying}}, \bibinfo {author} {\bibfnamefont {Z.-Z.}\ \bibnamefont {Zhang}},
  \bibinfo {author} {\bibfnamefont {J.}~\bibnamefont {Moser}}, \bibinfo
  {author} {\bibfnamefont {Z.-J.}\ \bibnamefont {Su}}, \bibinfo {author}
  {\bibfnamefont {X.-X.}\ \bibnamefont {Song}},\ and\ \bibinfo {author}
  {\bibfnamefont {G.-P.}\ \bibnamefont {Guo}},\ }\bibfield  {title} {\bibinfo
  {title} {Sliding nanomechanical resonators},\ }\bibfield  {journal} {\bibinfo
   {journal} {Nature Communications}\ }\textbf {\bibinfo {volume} {13}},\ \href
  {https://doi.org/10.1038/s41467-022-34144-5} {10.1038/s41467-022-34144-5}
  (\bibinfo {year} {2022})\BibitemShut {NoStop}%
\bibitem [{\citenamefont {Wang}\ \emph {et~al.}(2018)\citenamefont {Wang},
  \citenamefont {Zhu}, \citenamefont {Yang}, \citenamefont {Yuan},
  \citenamefont {Li}, \citenamefont {Wang}, \citenamefont {Chen}, \citenamefont
  {Deng}, \citenamefont {Liang}, \citenamefont {Li}, \citenamefont {Fan},
  \citenamefont {Guo},\ and\ \citenamefont {Jiang}}]{Wang2018-RN202}%
  \BibitemOpen
  \bibfield  {author} {\bibinfo {author} {\bibfnamefont {X.}~\bibnamefont
  {Wang}}, \bibinfo {author} {\bibfnamefont {D.}~\bibnamefont {Zhu}}, \bibinfo
  {author} {\bibfnamefont {X.}~\bibnamefont {Yang}}, \bibinfo {author}
  {\bibfnamefont {L.}~\bibnamefont {Yuan}}, \bibinfo {author} {\bibfnamefont
  {H.}~\bibnamefont {Li}}, \bibinfo {author} {\bibfnamefont {J.}~\bibnamefont
  {Wang}}, \bibinfo {author} {\bibfnamefont {M.}~\bibnamefont {Chen}}, \bibinfo
  {author} {\bibfnamefont {G.}~\bibnamefont {Deng}}, \bibinfo {author}
  {\bibfnamefont {W.}~\bibnamefont {Liang}}, \bibinfo {author} {\bibfnamefont
  {Q.}~\bibnamefont {Li}}, \bibinfo {author} {\bibfnamefont {S.}~\bibnamefont
  {Fan}}, \bibinfo {author} {\bibfnamefont {G.}~\bibnamefont {Guo}},\ and\
  \bibinfo {author} {\bibfnamefont {K.}~\bibnamefont {Jiang}},\ }\bibfield
  {title} {\bibinfo {title} {Stressed carbon nanotube devices for high
  tunability, high quality factor, single mode ghz resonators},\ }\href
  {https://doi.org/10.1007/s12274-018-2085-x} {\bibfield  {journal} {\bibinfo
  {journal} {Nano Research}\ }\textbf {\bibinfo {volume} {11}},\ \bibinfo
  {pages} {5812} (\bibinfo {year} {2018})}\BibitemShut {NoStop}%
\bibitem [{\citenamefont {Keşkekler}\ \emph {et~al.}(2021)\citenamefont
  {Keşkekler}, \citenamefont {Shoshani}, \citenamefont {Lee}, \citenamefont
  {Van Der~Zant}, \citenamefont {Steeneken},\ and\ \citenamefont
  {Alijani}}]{Keskekler2021-RN67}%
  \BibitemOpen
  \bibfield  {author} {\bibinfo {author} {\bibfnamefont {A.}~\bibnamefont
  {Keşkekler}}, \bibinfo {author} {\bibfnamefont {O.}~\bibnamefont
  {Shoshani}}, \bibinfo {author} {\bibfnamefont {M.}~\bibnamefont {Lee}},
  \bibinfo {author} {\bibfnamefont {H.~S.~J.}\ \bibnamefont {Van Der~Zant}},
  \bibinfo {author} {\bibfnamefont {P.~G.}\ \bibnamefont {Steeneken}},\ and\
  \bibinfo {author} {\bibfnamefont {F.}~\bibnamefont {Alijani}},\ }\bibfield
  {title} {\bibinfo {title} {Tuning nonlinear damping in graphene
  nanoresonators by parametric–direct internal resonance},\ }\bibfield
  {journal} {\bibinfo  {journal} {Nature Communications}\ }\textbf {\bibinfo
  {volume} {12}},\ \href {https://doi.org/10.1038/s41467-021-21334-w}
  {10.1038/s41467-021-21334-w} (\bibinfo {year} {2021})\BibitemShut {NoStop}%
\bibitem [{\citenamefont {Keşkekler}\ \emph {et~al.}(2023)\citenamefont
  {Keşkekler}, \citenamefont {Bos}, \citenamefont {Aragón}, \citenamefont
  {Alijani},\ and\ \citenamefont {Steeneken}}]{Keskekler2023-RN257}%
  \BibitemOpen
  \bibfield  {author} {\bibinfo {author} {\bibfnamefont {A.}~\bibnamefont
  {Keşkekler}}, \bibinfo {author} {\bibfnamefont {V.}~\bibnamefont {Bos}},
  \bibinfo {author} {\bibfnamefont {A.~M.}\ \bibnamefont {Aragón}}, \bibinfo
  {author} {\bibfnamefont {F.}~\bibnamefont {Alijani}},\ and\ \bibinfo {author}
  {\bibfnamefont {P.~G.}\ \bibnamefont {Steeneken}},\ }\bibfield  {title}
  {\bibinfo {title} {Multimode nonlinear dynamics of graphene resonators},\
  }\href {https://doi.org/10.1103/PhysRevApplied.20.064020} {\bibfield
  {journal} {\bibinfo  {journal} {Physical Review Applied}\ }\textbf {\bibinfo
  {volume} {20}},\ \bibinfo {pages} {064020} (\bibinfo {year}
  {2023})}\BibitemShut {NoStop}%
\bibitem [{\citenamefont {Zener}(1937)}]{Zener1937-RN217}%
  \BibitemOpen
  \bibfield  {author} {\bibinfo {author} {\bibfnamefont {C.}~\bibnamefont
  {Zener}},\ }\bibfield  {title} {\bibinfo {title} {Internal friction in
  solids. i. theory of internal friction in reeds},\ }\href
  {https://doi.org/10.1103/physrev.52.230} {\bibfield  {journal} {\bibinfo
  {journal} {Physical Review}\ }\textbf {\bibinfo {volume} {52}},\ \bibinfo
  {pages} {230} (\bibinfo {year} {1937})}\BibitemShut {NoStop}%
\bibitem [{\citenamefont {Schmid}\ \emph {et~al.}(2011)\citenamefont {Schmid},
  \citenamefont {Jensen}, \citenamefont {Nielsen},\ and\ \citenamefont
  {Boisen}}]{Schmid2011-RN194}%
  \BibitemOpen
  \bibfield  {author} {\bibinfo {author} {\bibfnamefont {S.}~\bibnamefont
  {Schmid}}, \bibinfo {author} {\bibfnamefont {K.~D.}\ \bibnamefont {Jensen}},
  \bibinfo {author} {\bibfnamefont {K.~H.}\ \bibnamefont {Nielsen}},\ and\
  \bibinfo {author} {\bibfnamefont {A.}~\bibnamefont {Boisen}},\ }\bibfield
  {title} {\bibinfo {title} {Damping mechanisms in high-q micro and
  nanomechanical string resonators},\ }\bibfield  {journal} {\bibinfo
  {journal} {Physical Review B}\ }\textbf {\bibinfo {volume} {84}},\ \href
  {https://doi.org/10.1103/physrevb.84.165307} {10.1103/physrevb.84.165307}
  (\bibinfo {year} {2011})\BibitemShut {NoStop}%
\bibitem [{\citenamefont {Nicholl}\ \emph {et~al.}(2017)\citenamefont
  {Nicholl}, \citenamefont {Lavrik}, \citenamefont {Vlassiouk}, \citenamefont
  {Srijanto},\ and\ \citenamefont {Bolotin}}]{Nicholl2017-RN118}%
  \BibitemOpen
  \bibfield  {author} {\bibinfo {author} {\bibfnamefont {R.~J.}\ \bibnamefont
  {Nicholl}}, \bibinfo {author} {\bibfnamefont {N.~V.}\ \bibnamefont {Lavrik}},
  \bibinfo {author} {\bibfnamefont {I.}~\bibnamefont {Vlassiouk}}, \bibinfo
  {author} {\bibfnamefont {B.~R.}\ \bibnamefont {Srijanto}},\ and\ \bibinfo
  {author} {\bibfnamefont {K.~I.}\ \bibnamefont {Bolotin}},\ }\bibfield
  {title} {\bibinfo {title} {Hidden area and mechanical nonlinearities in
  freestanding graphene},\ }\bibfield  {journal} {\bibinfo  {journal} {Physical
  Review Letters}\ }\textbf {\bibinfo {volume} {118}},\ \href
  {https://doi.org/10.1103/physrevlett.118.266101}
  {10.1103/physrevlett.118.266101} (\bibinfo {year} {2017})\BibitemShut
  {NoStop}%
\bibitem [{\citenamefont {Dolleman}\ \emph
  {et~al.}(2020{\natexlab{b}})\citenamefont {Dolleman}, \citenamefont
  {Blanter}, \citenamefont {Van Der~Zant}, \citenamefont {Steeneken},\ and\
  \citenamefont {Verbiest}}]{Dolleman2020-RN221}%
  \BibitemOpen
  \bibfield  {author} {\bibinfo {author} {\bibfnamefont {R.~J.}\ \bibnamefont
  {Dolleman}}, \bibinfo {author} {\bibfnamefont {Y.~M.}\ \bibnamefont
  {Blanter}}, \bibinfo {author} {\bibfnamefont {H.~S.~J.}\ \bibnamefont {Van
  Der~Zant}}, \bibinfo {author} {\bibfnamefont {P.~G.}\ \bibnamefont
  {Steeneken}},\ and\ \bibinfo {author} {\bibfnamefont {G.~J.}\ \bibnamefont
  {Verbiest}},\ }\bibfield  {title} {\bibinfo {title} {Phonon scattering at
  kinks in suspended graphene},\ }\bibfield  {journal} {\bibinfo  {journal}
  {Physical Review B}\ }\textbf {\bibinfo {volume} {101}},\ \href
  {https://doi.org/10.1103/physrevb.101.115411} {10.1103/physrevb.101.115411}
  (\bibinfo {year} {2020}{\natexlab{b}})\BibitemShut {NoStop}%
\bibitem [{\citenamefont {Steeneken}\ \emph {et~al.}(2013)\citenamefont
  {Steeneken}, \citenamefont {Ruigrok}, \citenamefont {Kang}, \citenamefont
  {Van~Beek}, \citenamefont {Bontemps},\ and\ \citenamefont
  {Koning}}]{Steeneken2013-RN216}%
  \BibitemOpen
  \bibfield  {author} {\bibinfo {author} {\bibfnamefont {P.}~\bibnamefont
  {Steeneken}}, \bibinfo {author} {\bibfnamefont {J.}~\bibnamefont {Ruigrok}},
  \bibinfo {author} {\bibfnamefont {S.}~\bibnamefont {Kang}}, \bibinfo {author}
  {\bibfnamefont {J.}~\bibnamefont {Van~Beek}}, \bibinfo {author}
  {\bibfnamefont {J.}~\bibnamefont {Bontemps}},\ and\ \bibinfo {author}
  {\bibfnamefont {J.-J.}\ \bibnamefont {Koning}},\ }\bibfield  {title}
  {\bibinfo {title} {Parameter extraction and support-loss in mems
  resonators},\ }\href@noop {} {\bibfield  {journal} {\bibinfo  {journal}
  {arXiv preprint arXiv:1304.7953}\ } (\bibinfo {year} {2013})}\BibitemShut
  {NoStop}%
\bibitem [{\citenamefont {Xu}\ \emph {et~al.}(2023)\citenamefont {Xu},
  \citenamefont {Shin}, \citenamefont {Sberna}, \citenamefont {Van Der~Kolk},
  \citenamefont {Cupertino}, \citenamefont {Bessa},\ and\ \citenamefont
  {Norte}}]{Xu2023-RN232}%
  \BibitemOpen
  \bibfield  {author} {\bibinfo {author} {\bibfnamefont {M.}~\bibnamefont
  {Xu}}, \bibinfo {author} {\bibfnamefont {D.}~\bibnamefont {Shin}}, \bibinfo
  {author} {\bibfnamefont {P.~M.}\ \bibnamefont {Sberna}}, \bibinfo {author}
  {\bibfnamefont {R.}~\bibnamefont {Van Der~Kolk}}, \bibinfo {author}
  {\bibfnamefont {A.}~\bibnamefont {Cupertino}}, \bibinfo {author}
  {\bibfnamefont {M.~A.}\ \bibnamefont {Bessa}},\ and\ \bibinfo {author}
  {\bibfnamefont {R.~A.}\ \bibnamefont {Norte}},\ }\bibfield  {title} {\bibinfo
  {title} {High‐strength amorphous silicon carbide for nanomechanics},\
  }\bibfield  {journal} {\bibinfo  {journal} {Advanced Materials}\ }\href
  {https://doi.org/10.1002/adma.202306513} {10.1002/adma.202306513} (\bibinfo
  {year} {2023})\BibitemShut {NoStop}%
\bibitem [{\citenamefont {Shin}\ \emph {et~al.}(2022)\citenamefont {Shin},
  \citenamefont {Cupertino}, \citenamefont {de~Jong}, \citenamefont
  {Steeneken}, \citenamefont {Bessa},\ and\ \citenamefont
  {Norte}}]{Shin2022-RN242}%
  \BibitemOpen
  \bibfield  {author} {\bibinfo {author} {\bibfnamefont {D.}~\bibnamefont
  {Shin}}, \bibinfo {author} {\bibfnamefont {A.}~\bibnamefont {Cupertino}},
  \bibinfo {author} {\bibfnamefont {M.~H.~J.}\ \bibnamefont {de~Jong}},
  \bibinfo {author} {\bibfnamefont {P.~G.}\ \bibnamefont {Steeneken}}, \bibinfo
  {author} {\bibfnamefont {M.~A.}\ \bibnamefont {Bessa}},\ and\ \bibinfo
  {author} {\bibfnamefont {R.~A.}\ \bibnamefont {Norte}},\ }\bibfield  {title}
  {\bibinfo {title} {Spiderweb nanomechanical resonators via bayesian
  optimization: Inspired by nature and guided by machine learning},\ }\href
  {https://doi.org/https://doi.org/10.1002/adma.202106248} {\bibfield
  {journal} {\bibinfo  {journal} {Advanced Materials}\ }\textbf {\bibinfo
  {volume} {34}},\ \bibinfo {pages} {2106248} (\bibinfo {year}
  {2022})}\BibitemShut {NoStop}%
\bibitem [{\citenamefont {Cupertino}\ \emph {et~al.}(2023)\citenamefont
  {Cupertino}, \citenamefont {Shin}, \citenamefont {Guo}, \citenamefont
  {Steeneken}, \citenamefont {Bessa},\ and\ \citenamefont
  {Norte}}]{Cupertino2023-RN243}%
  \BibitemOpen
  \bibfield  {author} {\bibinfo {author} {\bibfnamefont {A.}~\bibnamefont
  {Cupertino}}, \bibinfo {author} {\bibfnamefont {D.}~\bibnamefont {Shin}},
  \bibinfo {author} {\bibfnamefont {L.}~\bibnamefont {Guo}}, \bibinfo {author}
  {\bibfnamefont {P.~G.}\ \bibnamefont {Steeneken}}, \bibinfo {author}
  {\bibfnamefont {M.}~\bibnamefont {Bessa}},\ and\ \bibinfo {author}
  {\bibfnamefont {R.}~\bibnamefont {Norte}},\ }\href@noop {} {\emph {\bibinfo
  {title} {Centimeter-scale nanomechanical resonators with low dissipation}}}\
  (\bibinfo {year} {2023})\BibitemShut {NoStop}%
\bibitem [{\citenamefont {Lindahl}\ \emph {et~al.}(2012)\citenamefont
  {Lindahl}, \citenamefont {Midtvedt}, \citenamefont {Svensson}, \citenamefont
  {Nerushev}, \citenamefont {Lindvall}, \citenamefont {Isacsson},\ and\
  \citenamefont {Campbell}}]{Lindahl2012-RN237}%
  \BibitemOpen
  \bibfield  {author} {\bibinfo {author} {\bibfnamefont {N.}~\bibnamefont
  {Lindahl}}, \bibinfo {author} {\bibfnamefont {D.}~\bibnamefont {Midtvedt}},
  \bibinfo {author} {\bibfnamefont {J.}~\bibnamefont {Svensson}}, \bibinfo
  {author} {\bibfnamefont {O.~A.}\ \bibnamefont {Nerushev}}, \bibinfo {author}
  {\bibfnamefont {N.}~\bibnamefont {Lindvall}}, \bibinfo {author}
  {\bibfnamefont {A.}~\bibnamefont {Isacsson}},\ and\ \bibinfo {author}
  {\bibfnamefont {E.~E.~B.}\ \bibnamefont {Campbell}},\ }\bibfield  {title}
  {\bibinfo {title} {Determination of the bending rigidity of graphene via
  electrostatic actuation of buckled membranes},\ }\href
  {https://doi.org/10.1021/nl301080v} {\bibfield  {journal} {\bibinfo
  {journal} {Nano Letters}\ }\textbf {\bibinfo {volume} {12}},\ \bibinfo
  {pages} {3526} (\bibinfo {year} {2012})}\BibitemShut {NoStop}%
\bibitem [{\citenamefont {Goldsche}\ \emph
  {et~al.}(2018{\natexlab{b}})\citenamefont {Goldsche}, \citenamefont
  {Sonntag}, \citenamefont {Khodkov}, \citenamefont {Verbiest}, \citenamefont
  {Reichardt}, \citenamefont {Neumann}, \citenamefont {Ouaj}, \citenamefont
  {Von Den~Driesch}, \citenamefont {Buca},\ and\ \citenamefont
  {Stampfer}}]{Goldsche2018-RN14}%
  \BibitemOpen
  \bibfield  {author} {\bibinfo {author} {\bibfnamefont {M.}~\bibnamefont
  {Goldsche}}, \bibinfo {author} {\bibfnamefont {J.}~\bibnamefont {Sonntag}},
  \bibinfo {author} {\bibfnamefont {T.}~\bibnamefont {Khodkov}}, \bibinfo
  {author} {\bibfnamefont {G.~J.}\ \bibnamefont {Verbiest}}, \bibinfo {author}
  {\bibfnamefont {S.}~\bibnamefont {Reichardt}}, \bibinfo {author}
  {\bibfnamefont {C.}~\bibnamefont {Neumann}}, \bibinfo {author} {\bibfnamefont
  {T.}~\bibnamefont {Ouaj}}, \bibinfo {author} {\bibfnamefont {N.}~\bibnamefont
  {Von Den~Driesch}}, \bibinfo {author} {\bibfnamefont {D.}~\bibnamefont
  {Buca}},\ and\ \bibinfo {author} {\bibfnamefont {C.}~\bibnamefont
  {Stampfer}},\ }\bibfield  {title} {\bibinfo {title} {Tailoring mechanically
  tunable strain fields in graphene},\ }\href
  {https://doi.org/10.1021/acs.nanolett.7b04774} {\bibfield  {journal}
  {\bibinfo  {journal} {Nano Letters}\ }\textbf {\bibinfo {volume} {18}},\
  \bibinfo {pages} {1707} (\bibinfo {year} {2018}{\natexlab{b}})}\BibitemShut
  {NoStop}%
\bibitem [{\citenamefont {Villanueva}\ and\ \citenamefont
  {Schmid}(2014)}]{Villanueva2014-RN215}%
  \BibitemOpen
  \bibfield  {author} {\bibinfo {author} {\bibfnamefont {L.}~\bibnamefont
  {Villanueva}}\ and\ \bibinfo {author} {\bibfnamefont {S.}~\bibnamefont
  {Schmid}},\ }\bibfield  {title} {\bibinfo {title} {Evidence of surface loss
  as ubiquitous limiting damping mechanism in sin micro- and nanomechanical
  resonators},\ }\bibfield  {journal} {\bibinfo  {journal} {Physical Review
  Letters}\ }\textbf {\bibinfo {volume} {113}},\ \href
  {https://doi.org/10.1103/physrevlett.113.227201}
  {10.1103/physrevlett.113.227201} (\bibinfo {year} {2014})\BibitemShut
  {NoStop}%
\bibitem [{\citenamefont {Castellanos-Gomez}\ \emph {et~al.}(2014)\citenamefont
  {Castellanos-Gomez}, \citenamefont {Buscema}, \citenamefont {Molenaar},
  \citenamefont {Singh}, \citenamefont {Janssen}, \citenamefont {Van
  Der~Zant},\ and\ \citenamefont {Steele}}]{Castellanos-Gomez2014-RN172}%
  \BibitemOpen
  \bibfield  {author} {\bibinfo {author} {\bibfnamefont {A.}~\bibnamefont
  {Castellanos-Gomez}}, \bibinfo {author} {\bibfnamefont {M.}~\bibnamefont
  {Buscema}}, \bibinfo {author} {\bibfnamefont {R.}~\bibnamefont {Molenaar}},
  \bibinfo {author} {\bibfnamefont {V.}~\bibnamefont {Singh}}, \bibinfo
  {author} {\bibfnamefont {L.}~\bibnamefont {Janssen}}, \bibinfo {author}
  {\bibfnamefont {H.~S.~J.}\ \bibnamefont {Van Der~Zant}},\ and\ \bibinfo
  {author} {\bibfnamefont {G.~A.}\ \bibnamefont {Steele}},\ }\bibfield  {title}
  {\bibinfo {title} {Deterministic transfer of two-dimensional materials by
  all-dry viscoelastic stamping},\ }\href
  {https://doi.org/10.1088/2053-1583/1/1/011002} {\bibfield  {journal}
  {\bibinfo  {journal} {2D Materials}\ }\textbf {\bibinfo {volume} {1}},\
  \bibinfo {pages} {011002} (\bibinfo {year} {2014})}\BibitemShut {NoStop}%
\bibitem [{\citenamefont {Cartamil-Bueno}\ \emph {et~al.}(2017)\citenamefont
  {Cartamil-Bueno}, \citenamefont {Cavalieri}, \citenamefont {Wang},
  \citenamefont {Houri}, \citenamefont {Hofmann},\ and\ \citenamefont {Van
  Der~Zant}}]{Cartamil-Bueno2017-RN223}%
  \BibitemOpen
  \bibfield  {author} {\bibinfo {author} {\bibfnamefont {S.~J.}\ \bibnamefont
  {Cartamil-Bueno}}, \bibinfo {author} {\bibfnamefont {M.}~\bibnamefont
  {Cavalieri}}, \bibinfo {author} {\bibfnamefont {R.}~\bibnamefont {Wang}},
  \bibinfo {author} {\bibfnamefont {S.}~\bibnamefont {Houri}}, \bibinfo
  {author} {\bibfnamefont {S.}~\bibnamefont {Hofmann}},\ and\ \bibinfo {author}
  {\bibfnamefont {H.~S.~J.}\ \bibnamefont {Van Der~Zant}},\ }\bibfield  {title}
  {\bibinfo {title} {Mechanical characterization and cleaning of cvd
  single-layer h-bn resonators},\ }\bibfield  {journal} {\bibinfo  {journal}
  {npj 2D Materials and Applications}\ }\textbf {\bibinfo {volume} {1}},\ \href
  {https://doi.org/10.1038/s41699-017-0020-8} {10.1038/s41699-017-0020-8}
  (\bibinfo {year} {2017})\BibitemShut {NoStop}%
\bibitem [{\citenamefont {Dolleman}\ \emph {et~al.}(2019)\citenamefont
  {Dolleman}, \citenamefont {Hsu}, \citenamefont {Vollebregt}, \citenamefont
  {Sader}, \citenamefont {Van Der~Zant}, \citenamefont {Steeneken},\ and\
  \citenamefont {Ghatkesar}}]{Dolleman2019-RN224}%
  \BibitemOpen
  \bibfield  {author} {\bibinfo {author} {\bibfnamefont {R.~J.}\ \bibnamefont
  {Dolleman}}, \bibinfo {author} {\bibfnamefont {M.}~\bibnamefont {Hsu}},
  \bibinfo {author} {\bibfnamefont {S.}~\bibnamefont {Vollebregt}}, \bibinfo
  {author} {\bibfnamefont {J.~E.}\ \bibnamefont {Sader}}, \bibinfo {author}
  {\bibfnamefont {H.~S.~J.}\ \bibnamefont {Van Der~Zant}}, \bibinfo {author}
  {\bibfnamefont {P.~G.}\ \bibnamefont {Steeneken}},\ and\ \bibinfo {author}
  {\bibfnamefont {M.~K.}\ \bibnamefont {Ghatkesar}},\ }\bibfield  {title}
  {\bibinfo {title} {Mass measurement of graphene using quartz crystal
  microbalances},\ }\href {https://doi.org/10.1063/1.5111086} {\bibfield
  {journal} {\bibinfo  {journal} {Applied Physics Letters}\ }\textbf {\bibinfo
  {volume} {115}},\ \bibinfo {pages} {053102} (\bibinfo {year}
  {2019})}\BibitemShut {NoStop}%
\bibitem [{\citenamefont {Luinstra}\ and\ \citenamefont
  {Borchardt}(2011)}]{Luinstra2011-RN225}%
  \BibitemOpen
  \bibfield  {author} {\bibinfo {author} {\bibfnamefont {G.~A.}\ \bibnamefont
  {Luinstra}}\ and\ \bibinfo {author} {\bibfnamefont {E.}~\bibnamefont
  {Borchardt}},\ }\bibinfo {title} {Material properties of poly(propylene
  carbonates)},\ in\ \href {https://doi.org/10.1007/12_2011_126} {\emph
  {\bibinfo {booktitle} {Synthetic Biodegradable Polymers}}}\ (\bibinfo
  {publisher} {Springer Berlin Heidelberg},\ \bibinfo {year} {2011})\ pp.\
  \bibinfo {pages} {29--48}\BibitemShut {NoStop}%
\bibitem [{\citenamefont {Kumar}\ \emph {et~al.}(2013)\citenamefont {Kumar},
  \citenamefont {Kim},\ and\ \citenamefont {Yang}}]{Kumar2013-RN226}%
  \BibitemOpen
  \bibfield  {author} {\bibinfo {author} {\bibfnamefont {K.}~\bibnamefont
  {Kumar}}, \bibinfo {author} {\bibfnamefont {Y.-S.}\ \bibnamefont {Kim}},\
  and\ \bibinfo {author} {\bibfnamefont {E.-H.}\ \bibnamefont {Yang}},\
  }\bibfield  {title} {\bibinfo {title} {The influence of thermal annealing to
  remove polymeric residue on the electronic doping and morphological
  characteristics of graphene},\ }\href
  {https://doi.org/10.1016/j.carbon.2013.07.088} {\bibfield  {journal}
  {\bibinfo  {journal} {Carbon}\ }\textbf {\bibinfo {volume} {65}},\ \bibinfo
  {pages} {35} (\bibinfo {year} {2013})}\BibitemShut {NoStop}%
\bibitem [{\citenamefont {Lin}\ \emph {et~al.}(2012)\citenamefont {Lin},
  \citenamefont {Lu}, \citenamefont {Yeh}, \citenamefont {Jin}, \citenamefont
  {Suenaga},\ and\ \citenamefont {Chiu}}]{Lin2012-RN227}%
  \BibitemOpen
  \bibfield  {author} {\bibinfo {author} {\bibfnamefont {Y.-C.}\ \bibnamefont
  {Lin}}, \bibinfo {author} {\bibfnamefont {C.-C.}\ \bibnamefont {Lu}},
  \bibinfo {author} {\bibfnamefont {C.-H.}\ \bibnamefont {Yeh}}, \bibinfo
  {author} {\bibfnamefont {C.}~\bibnamefont {Jin}}, \bibinfo {author}
  {\bibfnamefont {K.}~\bibnamefont {Suenaga}},\ and\ \bibinfo {author}
  {\bibfnamefont {P.-W.}\ \bibnamefont {Chiu}},\ }\bibfield  {title} {\bibinfo
  {title} {Graphene annealing: How clean can it be?},\ }\href
  {https://doi.org/10.1021/nl203733r} {\bibfield  {journal} {\bibinfo
  {journal} {Nano Letters}\ }\textbf {\bibinfo {volume} {12}},\ \bibinfo
  {pages} {414} (\bibinfo {year} {2012})}\BibitemShut {NoStop}%
\bibitem [{\citenamefont {Hopcroft}\ \emph {et~al.}(2010)\citenamefont
  {Hopcroft}, \citenamefont {Nix},\ and\ \citenamefont
  {Kenny}}]{Hopcroft2010-RN181}%
  \BibitemOpen
  \bibfield  {author} {\bibinfo {author} {\bibfnamefont {M.~A.}\ \bibnamefont
  {Hopcroft}}, \bibinfo {author} {\bibfnamefont {W.~D.}\ \bibnamefont {Nix}},\
  and\ \bibinfo {author} {\bibfnamefont {T.~W.}\ \bibnamefont {Kenny}},\
  }\bibfield  {title} {\bibinfo {title} {What is the young's modulus of
  silicon?},\ }\href {https://doi.org/10.1109/jmems.2009.2039697} {\bibfield
  {journal} {\bibinfo  {journal} {Journal of Microelectromechanical Systems}\
  }\textbf {\bibinfo {volume} {19}},\ \bibinfo {pages} {229} (\bibinfo {year}
  {2010})}\BibitemShut {NoStop}%
\bibitem [{\citenamefont {Schmid}\ and\ \citenamefont
  {Hierold}(2008)}]{Schmid2008-RN220}%
  \BibitemOpen
  \bibfield  {author} {\bibinfo {author} {\bibfnamefont {S.}~\bibnamefont
  {Schmid}}\ and\ \bibinfo {author} {\bibfnamefont {C.}~\bibnamefont
  {Hierold}},\ }\bibfield  {title} {\bibinfo {title} {Damping mechanisms of
  single-clamped and prestressed double-clamped resonant polymer microbeams},\
  }\bibfield  {journal} {\bibinfo  {journal} {Journal of Applied Physics}\
  }\textbf {\bibinfo {volume} {104}},\ \href
  {https://doi.org/10.1063/1.3008032} {10.1063/1.3008032} (\bibinfo {year}
  {2008})\BibitemShut {NoStop}%
\bibitem [{\citenamefont {Castellanos-Gomez}\ \emph {et~al.}(2015)\citenamefont
  {Castellanos-Gomez}, \citenamefont {Singh}, \citenamefont {van~der Zant},\
  and\ \citenamefont {Steele}}]{Castellanoz-Gomez2015-RN49}%
  \BibitemOpen
  \bibfield  {author} {\bibinfo {author} {\bibfnamefont {A.}~\bibnamefont
  {Castellanos-Gomez}}, \bibinfo {author} {\bibfnamefont {V.}~\bibnamefont
  {Singh}}, \bibinfo {author} {\bibfnamefont {H.~S.~J.}\ \bibnamefont {van~der
  Zant}},\ and\ \bibinfo {author} {\bibfnamefont {G.~A.}\ \bibnamefont
  {Steele}},\ }\bibfield  {title} {\bibinfo {title} {Mechanics of
  freely-suspended ultrathin layered materials},\ }\href
  {https://doi.org/https://doi.org/10.1002/andp.201400153} {\bibfield
  {journal} {\bibinfo  {journal} {Annalen der Physik}\ }\textbf {\bibinfo
  {volume} {527}},\ \bibinfo {pages} {27} (\bibinfo {year} {2015})}\BibitemShut
  {NoStop}%
\bibitem [{\citenamefont {Castellanos‐Gomez}\ \emph
  {et~al.}(2013)\citenamefont {Castellanos‐Gomez}, \citenamefont
  {Van~Leeuwen}, \citenamefont {Buscema}, \citenamefont {Van Der~Zant},
  \citenamefont {Steele},\ and\ \citenamefont
  {Venstra}}]{Castellanos-Gomez2013-RN199}%
  \BibitemOpen
  \bibfield  {author} {\bibinfo {author} {\bibfnamefont {A.}~\bibnamefont
  {Castellanos‐Gomez}}, \bibinfo {author} {\bibfnamefont {R.}~\bibnamefont
  {Van~Leeuwen}}, \bibinfo {author} {\bibfnamefont {M.}~\bibnamefont
  {Buscema}}, \bibinfo {author} {\bibfnamefont {H.~S.~J.}\ \bibnamefont {Van
  Der~Zant}}, \bibinfo {author} {\bibfnamefont {G.~A.}\ \bibnamefont
  {Steele}},\ and\ \bibinfo {author} {\bibfnamefont {W.~J.}\ \bibnamefont
  {Venstra}},\ }\bibfield  {title} {\bibinfo {title} {Single‐layer mos2
  mechanical resonators},\ }\href {https://doi.org/10.1002/adma.201303569}
  {\bibfield  {journal} {\bibinfo  {journal} {Advanced Materials}\ }\textbf
  {\bibinfo {volume} {25}},\ \bibinfo {pages} {6719} (\bibinfo {year}
  {2013})}\BibitemShut {NoStop}%
\bibitem [{\citenamefont {Unterreithmeier}\ \emph {et~al.}(2010)\citenamefont
  {Unterreithmeier}, \citenamefont {Faust},\ and\ \citenamefont
  {Kotthaus}}]{Unterreithmeier20102010-RN192}%
  \BibitemOpen
  \bibfield  {author} {\bibinfo {author} {\bibfnamefont {Q.~P.}\ \bibnamefont
  {Unterreithmeier}}, \bibinfo {author} {\bibfnamefont {T.}~\bibnamefont
  {Faust}},\ and\ \bibinfo {author} {\bibfnamefont {J.~P.}\ \bibnamefont
  {Kotthaus}},\ }\bibfield  {title} {\bibinfo {title} {Damping of
  nanomechanical resonators},\ }\bibfield  {journal} {\bibinfo  {journal}
  {Physical Review Letters}\ }\textbf {\bibinfo {volume} {105}},\ \href
  {https://doi.org/10.1103/physrevlett.105.027205}
  {10.1103/physrevlett.105.027205} (\bibinfo {year} {2010})\BibitemShut
  {NoStop}%
\end{thebibliography}%


\newpage
\onecolumngrid
\newpage

\begin{centering}
{\huge Supporting Information:}
{\huge Tuning dissipation dilution in 2D material resonators by MEMS-induced tension}\\

\ \\
Michiel P.F. Wopereis$^1$, Niels Bouman$^1$, Satadal Dutta$^1$, Peter\\ G. Steeneken$^1$, Farbod Alijani$^1$, and Gerard J. Verbiest$^1$\\

\ \\
$^1$\emph{Department of Precision and Microsystems Engineering,} 
\emph{Delft University of Technology, Mekelweg 2, 2628 CD Delft, The Netherlands}
\end{centering}

\newpage
\section{S1. Fabrication of the device}
We fabricate the devices in several steps. First, the MEMS are manufactured by the commercially available XMB10 process. This is followed by the transfer of the membranes on the MEMS using a custom-made PDMS stamp with PPC film. Then, the devices are mounted on a PCB carrier and wire-bonded. Additional short-circuit bonds are added on the edge of the PCB carrier to prevent electrostatic charging. The next step involves the removal of the PPC film from the chip using annealing. Finally, the membranes are clamped using EBID, and the short-circuit bonds are removed. This section describes each step in more detail.

\subsection{1.1 Preperation}
\begin{figure}[h]
    \centering
    \includegraphics[width=0.6\textwidth]{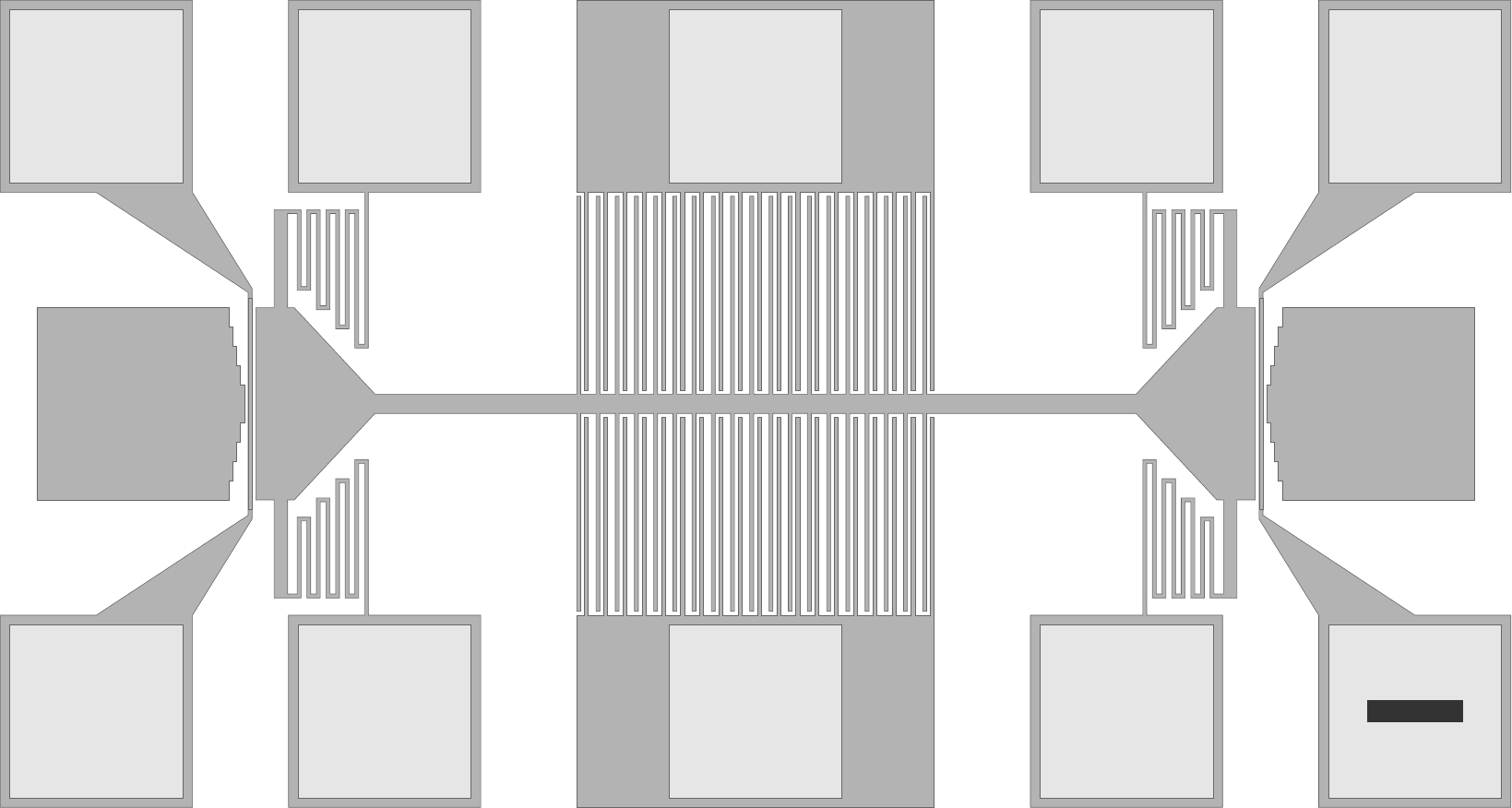}
    \caption{MEMS device layout. Scale bar: 50~\textmu m}
    \label{fig:mems-device-print}
\end{figure}

The membranes are strained in situ using MEMS devices. These devices are manufactured by the commercially available XMB10 process \cite{ZouYear-RN231, StreitYear-RN230}, which involves Deep Reactive Ion Etching (DRIE) to create a bottom cavity wafer with a gap of 50 \textmu m and a membrane wafer of 15 \textmu m. These wafers are then fusion-bonded with a 0.6 \textmu m $SiO_2$ in between the wafers to form the MEMS device (see Fig.~\ref{fig:mems-device-print}). 

\begin{figure}[h]
    \centering
    \includegraphics[width=1.0\textwidth]{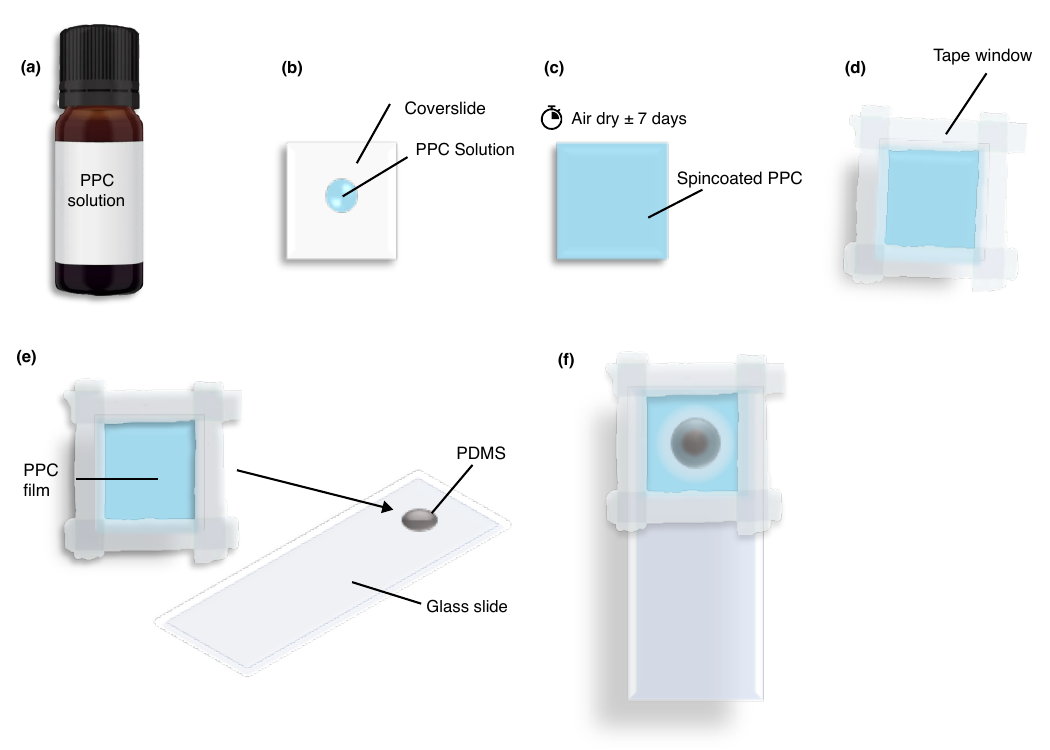}
    \caption{Preparation of the stamp for the transfer procedure. a) Solution of 15/85 mass ratio PPC granulate to anisole solvent b) Droplet of the PPC solution on a microscope cover slide c) Even distribution of the PPC solution over the cover slide after spin-coating d) Tape window attached to the PPC layer of the cover slide e) Transferring the PPC film to the dome-shaped stamp f) PDMS-dome shaped stamp covered with a PPC film ready for stamping.}
    \label{fig:attaching-ppc-to-pdms-stamp}
\end{figure}

\noindent Next, we produce a PPC film by dissolving PPC pellets \\  \ce{([CH(CH3)CH2OCO2]_n)} in anisole \ce{(C7H8O)} with a 15/85 ratio. This dissolution process is achieved using a magnetic stirrer at 50 degrees Celsius for 2 hours. Subsequently, microscope cover slides are cleaned with isopropanol, and a droplet of the PPC solution is carefully placed at the center of each cover slide. To ensure even coating, a spin coater is used to uniformly spread a thin layer of the PPC solution over the entire surface of the cover slide (see Figures \ref{fig:attaching-ppc-to-pdms-stamp}(a)-(d)). The slides are left to air-dry for seven days to complete the PPC film preparation, allowing complete evaporation of the anisole solvent. Finally, a custom dome-shaped PDMS stamp is produced for precise pick-and-place of the 2D material membranes. A mixture of 87\% PDMS resin and 13\% curing agent is mixed and dispensed as a droplet onto a microscope slide to make the stamp. The slide is then inverted and placed into a vacuum chamber for one hour to remove microscopic air bubbles. Afterward, the stamp is left to air-dry for 24 hours, forming a small dome on the surface. The final step is to cover the PDMS stamp with the PPC film, illustrated in Figure \ref{fig:attaching-ppc-to-pdms-stamp}.

\subsection*{1.2 Transfer of 2D Materials on MEMS}
\begin{figure}[h]
    \centering
    \includegraphics[width=1.0\textwidth]{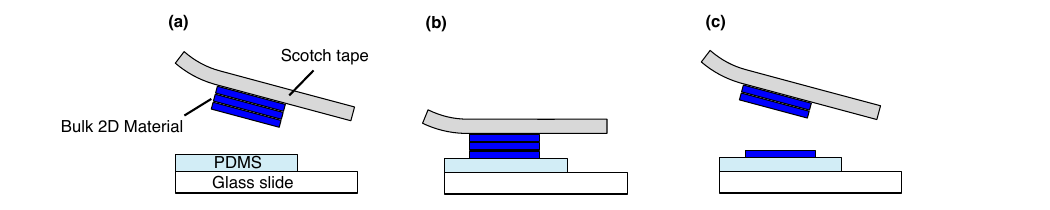}
    \caption{Schematic representation of the mechanical exfoliation of 2D materials on a PDMS substrate.}
    \label{fig:schematic-representation-of-mechanical-exfoliationl}
\end{figure}

Transferring 2D materials onto the suspended shuttle of the MEMS poses significant challenges. Conventional transfer methods, including viscoelastic stamping techniques \cite{Castellanos-Gomez2014-RN172}, have proven unsuitable for this complex task. To address this limitation, we present a novel transfer method utilizing a sacrificial PPC layer, later removed through an annealing process. In our approach, rectangular-shaped flakes were selected through traditional exfoliation methods \cite{Novoselov2005-RN222} and placed onto a rectangular square of polydimethylsiloxane \ce{CH3nSi(CH3)3} (PDMS) (see Figure \ref{fig:schematic-representation-of-mechanical-exfoliationl}). PDMS was chosen based on its transparency and adhesive properties.

\begin{figure}[h]
    \centering
    \includegraphics[width=1.0\textwidth]{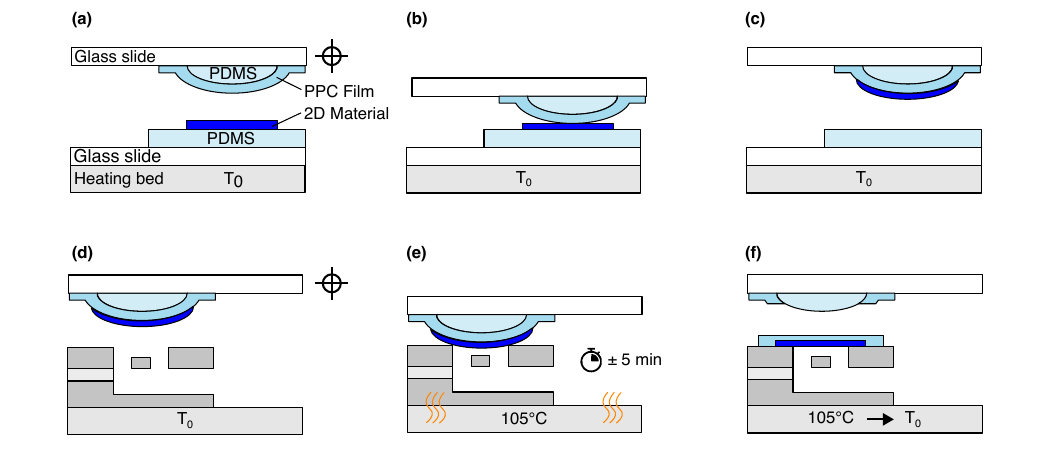}
    \caption{Procedure for transferring a 2D membrane to a substrate using a sacrificial PPC layer. a) Positioning the stamp above the membrane of interest b) Lower the stamp such that the membrane adheres to the PPC film of the stamp c) The membrane is removed from the PDMS d) Position the stamp above the cavity e) Making contact with the MEMS for about 5 minutes at 105$\degree C$ until the PPC layer starts to melt onto the MEMS f) Removing the stamp from the MEMS and cooling it down to room temperature.}
    \label{fig:transferring-2d-material-onto-mems}
\end{figure}

Next, we used a dome-shaped stamp, which was prepared in the first stage to transfer the membranes. This stamp was carefully positioned above the desired flake on the PDMS substrate (see Figure \ref{fig:transferring-2d-material-onto-mems}). By gently bringing the stamp with the PPC layer in contact with the 2D material, we exploited the higher adhesion of PPC in contrast with the PDMS substrate, which caused the flake to adhere to the stamp with the PPC layer (see Figure \ref{fig:transferring-2d-material-onto-mems}(b)-(c)). This approach is similar to the technique introduced by Kinoshita \cite{Kinoshita2019-RN41}.

\begin{figure}[h]
    \centering
    \includegraphics[width=1.0\textwidth]{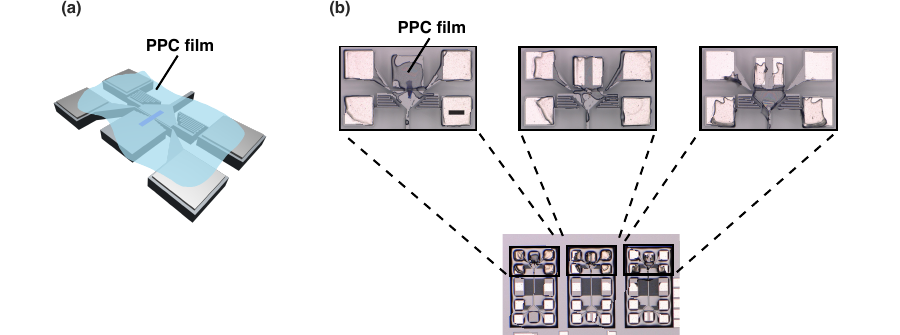}
    \caption{Device after transfer of 2D materials using a sacrificial PPC layer. a) Illustration of the device with the 2D material membrane beneath the PPC film b) Devices after transfer of three membranes onto a single die. Scale bar: 10 \textmu m }
    \label{fig:S1.2-defice-after-ppc-transfer}
\end{figure}

Next, the dome-shaped stamp with the 2D material was carefully positioned above the desired location on the MEMS and brought into contact. The MEMS device was then heated to 105$\degree C$, maintaining contact for approximately 5 minutes until the PPC layer started to melt (see Figures \ref{fig:transferring-2d-material-onto-mems}(d)-(e)). Afterward, the stamp was gradually removed, leaving the sacrificial PPC layer adhered to the MEMS device with the 2D material beneath it. The device was cooled to room temperature, completing the transfer procedure (see Figure \ref{fig:transferring-2d-material-onto-mems}f). An example of a die with MEMS after the membrane transfer can be found in Figure \ref{fig:S1.2-defice-after-ppc-transfer}b.

The presented method offers several distinct advantages. Notably, it allows for precise pick and place of flakes within 350 $\mu m$ diameter, enabling accurate and controlled placement of multiple flakes onto a single chip (Figure \ref{fig:S1.2-defice-after-ppc-transfer}b). Furthermore, the sacrificial PPC layer effectively accommodates transfer onto suspended and delicate MEMS components. Moreover, this transfer method is entirely dry, eliminating the need for wet chemistry commonly used with fragile parts \cite{Xie2021-RN59, Verbiest2021-RN15, SonntagYear-RN122}. This is especially beneficial for parts susceptible to surface tension forces, such as the comb drive. 

\subsection*{1.3 Carrier Bonding Procedure}
\begin{figure}[h]
    \centering
    \includegraphics[width=1.0\textwidth]{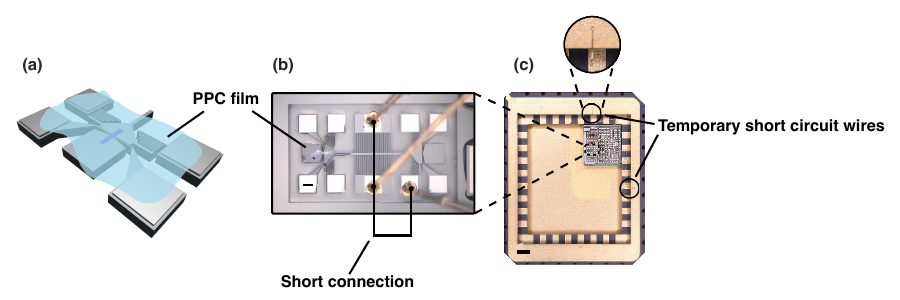}
    \caption{Device connected to a PCB carrier after wire bonding a) Illustration of the device with PPC film preventing the membrane from slippage during wire bonding b) Detailed view of wire-bonded MEMS. Scale bar: 50 \textmu m c) Silicon die containing MEMS wire bonded on a PCB carrier, including a detailed view of the short connection wires. Scale bar: 1000 \textmu m }
    \label{fig:S1.3-after-wire-bonding}
\end{figure}

The silicon die containing the MEMS is secured onto a chip carrier using a combination of silver paste and a droplet of super glue applied on one side of the chip. This gluing ensures a secure die attachment to the PCB carrier, essential for the wire bonding (see Figure \ref{fig:S1.3-after-wire-bonding}c). Subsequently, wire bonding is performed to establish electrical connections between the MEMS and the carrier chip. A TPT HB05 wire bonder is used for this purpose, using a 25\textmu m golden wire. 
For the initial ball bond, we select US, Time, and Force, 100, 200, and 10, respectively. For the second wedge bond, we select US, Time, and Force, 125, 200, and 15, respectively. Also, the stage is heated to 120 $\degree$C before bonding. 
During the bonding process, the PPC film plays a crucial role in effectively holding the 2D material in place, preventing the membrane from slipping (see Figure \ref{fig:S1.3-after-wire-bonding}(a)-(b)). As a precautionary measure, both sides of the comb drive fingers are short-circuited to prevent charges from accumulating over the comb drive, which could cause unwanted movements during fabrication. The short connection is created with two extra wire bonds to the edge of the PCB carrier (see detail view in Figure \ref{fig:S1.3-after-wire-bonding}c).

\subsection*{1.4 Annealing}
\begin{figure}[h]
    \centering
    \includegraphics[width=1.0\textwidth]{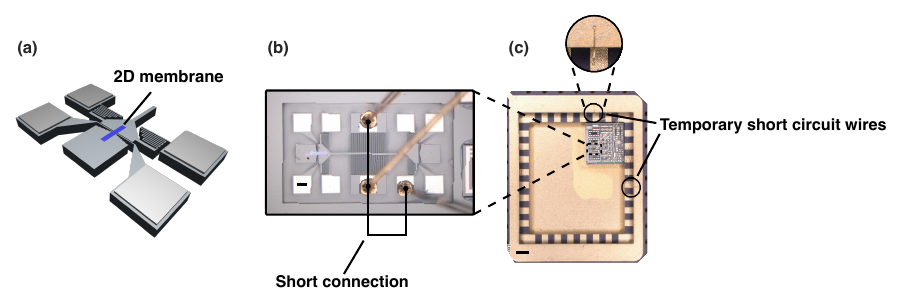}
    \caption{Device after the annealing process. a) Illustration of the device after annealing (the sacrificial PPC layer is evaporated) b) The wire-bonded MEMS after annealing. Scale bar: 10 \textmu m c) The silicon die including the MEMS attached to the PCB carrier after annealing for 3 hours at 300$\degree C$. Scale bar: 1000 \textmu m}
    \label{fig:S1.4-device-after-annealing}
\end{figure}

In the mechanical transfer step, the 2D material is placed onto the cavity of the MEMS using a sacrificial PPC film. The annealing step ensures that the PPC residue is removed from the MEMS. During annealing, the chip mounted on the PCB carrier is inserted into a high vacuum oven operating at a pressure below $10^{-5}$ mbar. The annealing is conducted at $300\degree C$ for 3 hours to ensure that all polymer contamination from the transfer is removed \cite{Cartamil-Bueno2017-RN223, Dolleman2019-RN224}. Notably, this annealing temperature exceeds the highest reported thermal decomposition temperature of PPC of $278 \degree C$\cite{Luinstra2011-RN225}, effectively removing any residual PPC while being significantly below the thermal thresholds of other materials on the chip. To prevent unwanted reactions, the vacuum oven is thoroughly flushed with Argon before the annealing process. This creates an oxygen-free environment, preventing potential interaction between PPC and oxygen molecules during annealing \cite{Kumar2013-RN226, Lin2012-RN227}. Figures \ref{fig:S1.4-device-after-annealing}(a)-(c) show the removal of the PPC film after annealing.

\subsection*{1.5 Electron beam induced deposition (EBID)}
\begin{figure}[h]
    \centering
    \includegraphics[width=1.0\textwidth]{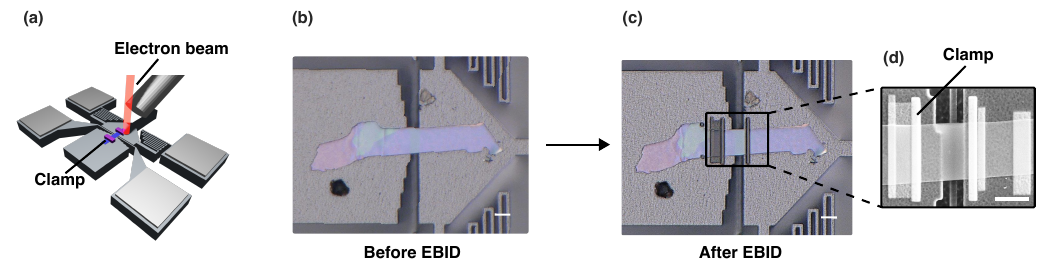}
    \caption{Electron beam induced deposition (EBID) procedure for clamping the membrane. a) Illustration of EBID b) Optical image of device D5 before EBID. Scale bar: 10 \textmu m. c) Optical image of the device after EBID. Scale bar: 10 \textmu m. d) SEM detailed view of the platinum clamps holding the membrane of device D5 in place. Scale bar: 10 \textmu m. }
    \label{fig:S1.5-Device-ebid-procedure}
\end{figure}

To prevent slippage of the 2D material when subjected to tensile forces surpassing the Van der Waals adhesion forces during the experiment, a protective layer of approximately 130nm thick platinum is carefully deposited on both ends of the 2D material. This deposition is conducted through electron beam-induced deposition (EBID) to ensure precise control over the added layer. For this purpose, the FEI SEM Helios G4 CX system with a gas injection system (GIS) is used, operating at 10kV and 11nA. We deposited the platinum layer with an offset from the suspended area to avoid contaminating the membranes during EBID. Refer to Figure \ref{fig:S1.5-Device-ebid-procedure} for an illustration. Applying this platinum protective layer provides a secure and reliable anchoring mechanism. The deposited platinum layer is a robust measure to firmly hold the 2D material, effectively preventing any undesirable slippage during the experiment. This essential precaution ensures the repeatability and accuracy of the experimental results, allowing for confident analysis of the 2D material under controlled conditions. In Figure \ref{fig:S1.5-Device-ebid-procedure}b and Figures (c)-(d), the device is shown before and after EBID, respectively.

\subsection*{1.6 Removing short circuit wires}
Prior to testing, the short connection added to prevent charges from accumulating over the comb drive during electron beam exposure needs to be removed (Figures \ref{fig:S1.4-device-after-annealing}(b)-(c)). These additional bonds placed on the PCB carrier’s edge are carefully removed under a microscope using a thin tungsten needle and tweezers.

\newpage
\section{S2. MEMS stiffness characterization}
\subsubsection{2.1 Back-of-the-envelope calculation}
\label{sec:back-of-the-envelope calculation device stiffness}
\begin{figure}[h]
    \centering
    \includegraphics[width=0.8\textwidth]{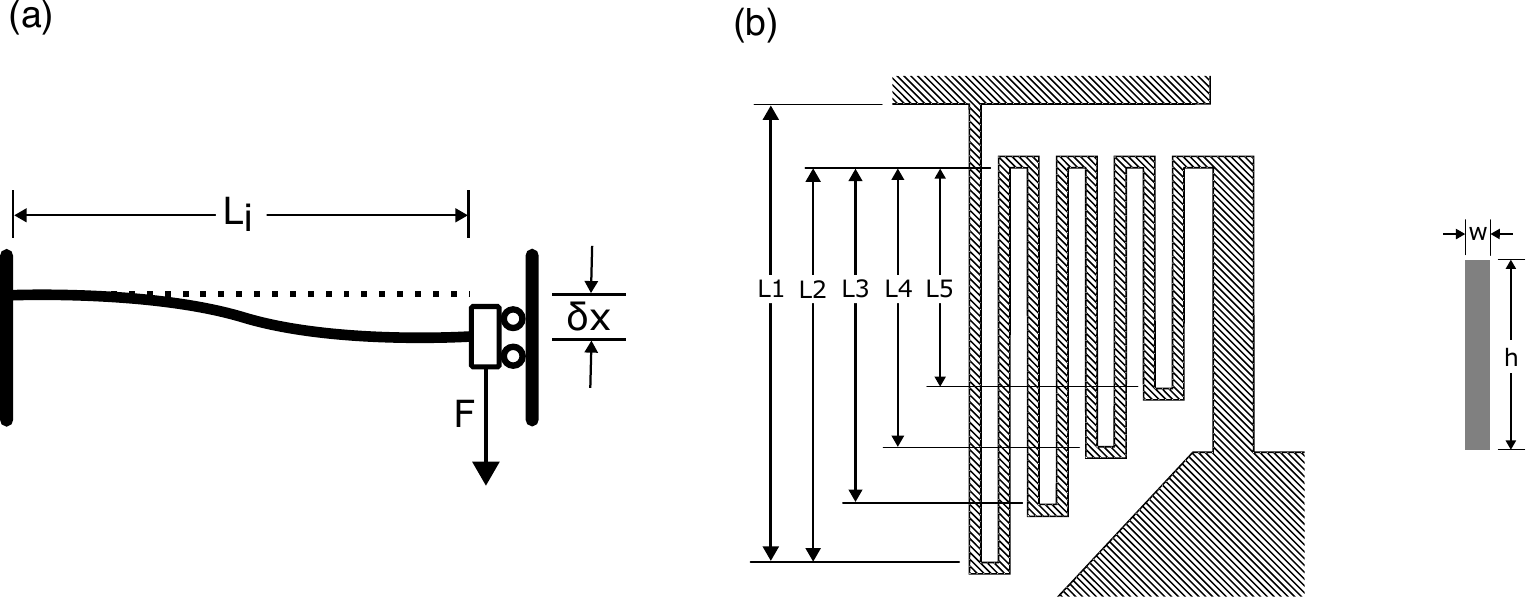}
    \caption{(a) Schematic representation of a single bent beam (b) Dimensions of non-uniform serpentine flexure}
    \label{fig:stiffness-of-non-uniform-serpentine-flexure}
\end{figure}

To accurately determine the in-plane displacement of the shuttle, it is essential to know the system's stiffness. Four non-uniform serpentine flexures support the shuttle of the comb drive. The serpentine flexure has the following dimensions: one beam of $L_1 = 79 \mu m$, one beam of $L_2 = 68 \mu m$, two beams of $L_3 = 58 \mu m$, two beams of $L_4 = 48 \mu m$ and two beams of $L_5 = 38 \mu m$ (see Figure \ref{fig:stiffness-of-non-uniform-serpentine-flexure}b). Each beam has a width (w) of 2$\mu m$ and a height (h) of 15$\mu m$. The Young's modulus (E) of the silicon flexure is 169GPa \cite{Hopcroft2010-RN181}. 
\noindent The stiffness of a single non-uniform serpentine flexure can be approximated by separating the flexure in a series of beams and calculating the stiffness of each beam individually, see Figure \ref{fig:stiffness-of-non-uniform-serpentine-flexure} and Equation \ref{eqn:stiffness-ks-calc}. 
\begin{equation}
\label{eqn:stiffness-ks-calc}
\begin{aligned}
& \delta x_i=\frac{F L_i^3}{12 E \cdot I} \\
& I=\frac{h  w^3}{12} \\
& k_i =\frac{F}{\delta x_i} \\
& k_i=E h \frac{w^3}{L_i^3} \\
\end{aligned}
\end{equation}
Then, the stiffness of the flexure can be estimated by summating the individual beam stiffnesses as a series of springs ($k_1$ to $k_5$). This neglects the contribution of the connecting side beams, which is assumed to be low. Following Equation \ref{eqn:stiffness-kcd-calc}, $k_s$ is calculated to be  13.3 N/m. Since the system consists of four flexures in parallel. The stiffness of the comb drive is four times the stiffness of a single flexure; thus, $k_\mathrm{cd} = 53.1 N/m$.
\begin{equation}
\label{eqn:stiffness-kcd-calc}
\begin{aligned}
& k_s \approx\left(\frac{2}{k_1}+\frac{2}{k_2}+\frac{2}{k_3}+\frac{1}{k_4}+\frac{1}{k_5}\right)^{-1} \\
& k_\mathrm{cd} = 4 k_s
\end{aligned}
\end{equation}
\subsubsection{2.2 Computational study of comb drive stiffness}
\begin{figure}[h]
  \begin{minipage}{1.0\textwidth} 
      \centering
      \begin{subfigure}[t]{0.6\textwidth}
        \centering
        \includegraphics[width=\textwidth]{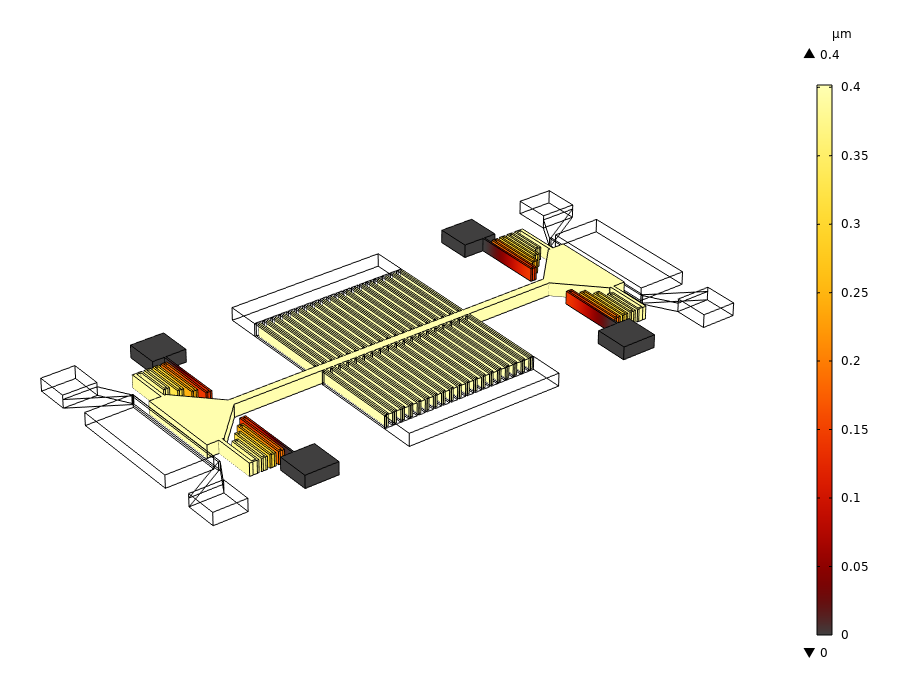}
        \caption{}
        \label{fig:comsol-stiffness-calc}
      \end{subfigure}
      \hfill
      \begin{subfigure}[t]{0.3\textwidth}
        \centering
        \includegraphics[width=\textwidth]{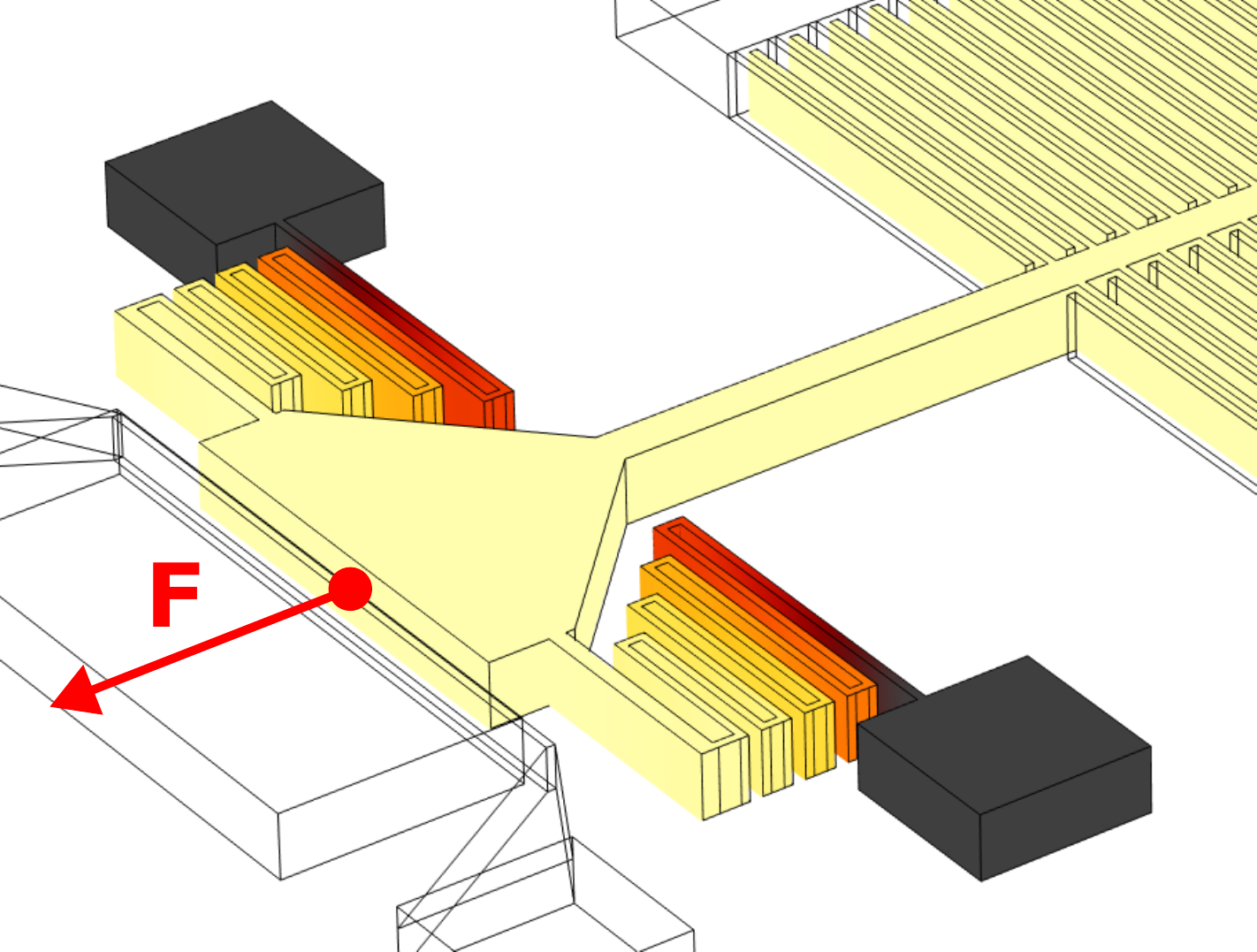}
        \caption{}
        \label{fig:comsol-stiffness-loads}
      \end{subfigure}
      \caption{Comsol Multiphysics study (a) Shuttle displacement with an applied load of $F=20 \mu N$ (b) Direction of the applied load on the shuttle}
      \label{fig:comsol-study}
    \end{minipage}
\end{figure}
A computational study was conducted using COMSOL Multiphysics. Within this analysis, a force of 20 $\mu N$ (F) was exerted at the endpoint of the shuttle, directed along the axis of the narrowest gap of the comb drive, see Figure \ref{fig:comsol-stiffness-loads}. This force induced a displacement of 0.4 $\mu m$ ($\Delta x$) in the shuttle's position (see Figure \ref{fig:comsol-stiffness-calc}). By employing the relation $k_\mathrm{cd} = F/\Delta x$, the inherent stiffness of the system can be determined, yielding $k_\mathrm{cd} = 50 N/m$. Notably, the determined stiffness value is marginally lower than the estimate from initial back-of-the-envelope calculations. This difference arises from including the whole flexure geometry in the numerical simulation, including the interconnecting beams. This factor contributes to a minor reduction in the overall stiffness of the device.

\subsubsection{2.3 Determining comb drive stiffness experimentally based on pull-in voltage}
When voltage is applied over a capacitance, the electrostatic force will work to reduce the gap between opposing plates. At small voltages, the electrostatic force is counterbalanced with the spring force; however, when the voltages are increased, the plates will eventually pull in. Since this specific pull-in voltage is determined by the stiffness of the springs of the comb drive, it is an effective way to experimentally determine the stiffness of the comb drive based on the pull-in voltage for an empty device. The force acting on the movable shuttle can be derived by Equation \ref{eqn:F_sys}. 

\begin{equation}
\label{eqn:F_sys}
    F_\mathrm{sys} = F_{cd}-k_\mathrm{cd} x = \frac{A \epsilon_0}{2}\left(\frac{1}{{{\left( d_1-x \right)^2}}}-\frac{1}{{{\left( d_2+x \right)^2}}} \right)V^2-k_\mathrm{cd} x
\end{equation}
When the system is in equilibrium, the electrostatic and spring forces cancel each other. The shuttle is stationary in this case and $F_\mathrm{sys}=0$. This leads to equation \ref{eqn:F_sys_equillibrium}. This equation can be used to calculate the shuttle position x as a function of voltage. Above the pull-in voltage ($V_{PI}$), Equation \ref{eqn:F_sys_equillibrium} has no solutions. The pull-in point can be obtained by examining the stiffness of the system, which can be obtained by Equation \ref{eqn:k_sys}.
\begin{equation}
\label{eqn:F_sys_equillibrium}
    \frac{A \epsilon_0}{2}\left(\frac{1}{{{\left( d_1-x \right)^2}}}-\frac{1}{{{\left( d_2+x \right)^2}}} \right)V^2 = k_\mathrm{cd} x
\end{equation}
\begin{equation}
\label{eqn:k_sys}
    \frac{\partial F_\mathrm{sys}}{\partial x} = A \epsilon_0\left(\frac{1}{{{\left( d_1-x \right)^3}}}-\frac{1}{{{\left( d_2+x \right)^3}}} \right)V^2 - k_\mathrm{cd} 
\end{equation}
When there is no voltage applied to the system, Equation \ref{eqn:k_sys} is $\frac{\partial F_\mathrm{sys}}{\partial x}=-k_\mathrm{cd}$; however, when V is increased, the stiffness will become less negative. At the pull-in point, $\frac{\partial F_\mathrm{sys}}{\partial x}=0$; hence, this point can be used to calculate the pull-in voltage for a certain stiffness $k_\mathrm{cd}$. The stiffness around the equilibrium point can be obtained by first solving Equation \ref{eqn:F_sys_equillibrium} for $V^2$, and then plugging in this solution into Equation \ref{eqn:k_sys}. Eventually, when setting $\frac{\partial F_\mathrm{sys}}{\partial x}=0$, the equation can be solved for the shuttle displacement x, hereby obtaining the position where pull-in occurs $x=x_\mathrm{PI}$:
\begin{equation}
\label{eqn:pull-in-discplacement_x_pi}
    x_\mathrm{PI} = \frac{\alpha ^2 + 3 \, \alpha \, d_1 - 3 \, \alpha \, d_2 + {\left(d_{1}-d_{2}\right)}^2}{8\,\alpha }
\end{equation}
with:
\begin{equation*}
    \alpha = {\left(4\,\sqrt{{\left({d_{1}}^2-{d_{2}}^2\right)}^2\,\left(5\,{d_{1}}^2+6\,d_{1}\,d_{2}+5\,{d_{2}}^2\right)}+5\,d_{1}\,{d_{2}}^2-5\,{d_{1}}^2\,d_{2}-9\,{d_{1}}^3+9\,{d_{2}}^3\right)}^{1/3}
\end{equation*}
Finally, knowing the unstable point $x_L$ purely based on the spacing of the comb fingers, one can determine either the pull-in point for a given $k_\mathrm{cd}$ with Equation \ref{eqn:k_cd_based_on_pull_in_voltage} or the stiffness of the comb drive with equation \ref{eqn:pull-in-voltage}.
\begin{equation}
\label{eqn:k_cd_based_on_pull_in_voltage}
    k_\mathrm{cd}=\frac{A \epsilon_0}{2 x}\left(\frac{1}{{{\left( d_1-x_\mathrm{PI} \right)^2}}}-\frac{1}{{{\left( d_2+x_\mathrm{PI} \right)^2}}} \right)V_\mathrm{PI}^2
\end{equation}
\begin{equation}
\label{eqn:pull-in-voltage}
    V_\mathrm{PI}=\sqrt{\frac{A \epsilon_0}{2 x k_\mathrm{cd}}\left(\frac{1}{{{\left( d_1-x_\mathrm{PI} \right)^2}}}-\frac{1}{{{\left( d_2+x_\mathrm{PI} \right)^2}}} \right)}
\end{equation}
\textbf{Results}
The pull-in test has been conducted for devices C8M1 and C8M4; both devices have a completely torn flake, so there is no stiffness contribution to the membrane. For the proceeding calculation of the devices that are not experimentally tested on stiffness, the average of devices C8M1 and C8M4 will be used (see Table~\ref{tab:experimental-stiffness-kcd}), which is $38.42$ N/m. The pull-in voltage is approximately 13.5 $\pm$ 0.5 V.

\begin{table}[]
\centering
\begin{tabular}{lll}
\hline
Device & \text{Voltage} & \text{Stiffness} \\ \hline
C8M1 & $13.89 \pm 0.04$ V & $41.07 \pm 0.23$ Nm$^{-1}$ \\ 
C8M4 & $12.96 \pm 0.06$ V & $35.76 \pm 0.33$ Nm$^{-1}$ \\ 
\end{tabular}
\caption{Experimental stiffness characterization}
\label{tab:experimental-stiffness-kcd}
\end{table}

\newpage
\section{S3. Dissipation dilution model}
Membranes can build up much potential energy when the vibrational deflection has to work against the high lateral tensile stress \cite{Schmid2016-RN203}, leading to dissipation dilution. The tensile stress of the membranes is incrementally increased using a comb drive actuator. By definition, the quality factor of a membrane subjected to tensile stress can be described by \cite{Schmid2011-RN194, Schmid2008-RN220}:

\begin{equation}
Q=2 \pi \frac{W_\mathrm{tension }+W_\mathrm{elongation}+W_\mathrm{bending}}{\Delta W_\mathrm{elongation}+\Delta W_\mathrm{bending}}
\end{equation}

\noindent Where $W_\mathrm{tension}$ is the stored elastic energy required to deflect the membrane against the tensile force, $W_\mathrm{elongation}$ and $W_\mathrm{bending}$ is the stored energy due to elongation and bending, respectively. Finally, $\Delta W_\mathrm{elongation}$ and $\Delta W_\mathrm{bending}$ are the lost energy due to elongation and bending. When the stored tensile energy dominates the mechanical behavior for highly stressed membranes, the elongation and bending energies become negligible ($W_\mathrm{tensile} \gg W_\mathrm{elongation} + W_\mathrm{bending}$). Moreover, when assumed that the intrinsic damping is equal for both the elongation and bending, such that $Q_{\mathrm{elongation }}=2 \pi \frac{W_\mathrm{clongation}}{\Delta W_\mathrm{elongation}}=Q_\mathrm{bending }=2 \pi \frac{W_\mathrm{bending}}{\Delta W_\mathrm{bending}}$, we can derive the expression for a highly stressed membrane as $Q \approx \alpha_\mathrm{dd} \cdot Q_\mathrm{intrinsic}$, where $\alpha_{dd}$ is the dilution factor, namely:

\begin{equation}
\alpha_{d d}=\left[\frac{W_\mathrm{bending}}{W_\mathrm{tensile}}+\frac{W_\mathrm{elongation}}{W_\mathrm{tensile}}\right]^{-1} .
\end{equation}

\noindent When the membranes are strained, the stored tensile energy $W_\mathrm{tensile}$ in the resonator increases while the energy stored in bending and elongation remains identical.\cite{Schmid2016-RN203} Therefore, the damping dilution factor becomes larger with increased strain. This increase in the stored tensile energy ‘dilutes’ the intrinsic losses $Q_{int}$, resulting in a higher $Q$. 

For unstrained membranes, the resonance frequency depends on the thickness of the flake. For very thin membranes ($\approx$ 1 to 5 layers), the resonance frequency is dominated by the initial pre-tension in the membrane $f_\mathrm{mem}$ \cite{Castellanoz-Gomez2015-RN49}. In contrast, relatively thick membranes ($> \approx$ 15 layers) follow the expected dynamics for plates $f_\mathrm{plate}$, which depend mainly on the geometry of the membranes \cite{Castellanoz-Gomez2015-RN49}. In the cross-over regime, between pre-tension dominated and bending rigidity dominated, the fundamental resonance frequency of a membrane subjected to strain can be estimated by $f_0 = \sqrt{f_\mathrm{mem}^2 + f_\mathrm{plate}^2}$ \cite{Castellanos-Gomez2013-RN199, Castellanoz-Gomez2015-RN49}, and the quality factor by due to dissipation dilution by\cite{Steeneken2021-RN1}:

\begin{equation}
\label{eqn:Q-eqn-Steeneken2021}
Q_D \approx\left(\frac{\left|f_{\mathrm{mem }}\right|^2}{\left|f_{\mathrm{plate }}\right|^2}+1\right) \frac{E_1}{E_2},
\end{equation}

\noindent Where $E_1$ and $E_2$ represent the dynamic modulus at a certain frequency by $E=E_1 + iE_2$ \cite{Schmid2016-RN203, Unterreithmeier20102010-RN192, Schmid2008-RN220}, with $E_1$ being the storage modulus and $E_2$ being the loss modulus. Assuming that $E_1$ and $E_2$ do not change much with frequency, we can approximate $E_1/E_2 \approx Q_\mathrm{int}$. Since the membranes are relatively thick and dominated by the bending rigidity, and $f_\mathrm{plate}$ does not change with strain, we can assume that $f_0$ equals $f_\mathrm{plate}$ when the membranes are unstrained ($V_\mathrm{cd}=0\, V$). Therefore, $f_\mathrm{plate} = f_0$(0 V). The dynamics can be described by $f_0=\sqrt{f_\mathrm{mem}^2 + f_\mathrm{plate}^2}$. This simplifies Equation \ref{eqn:Q-eqn-Steeneken2021} to:

\begin{equation}
    \label{eqn:Q-approximation}
    Q_D \approx \left(\frac{f_0}{f_\mathrm{plate}}\right)^2 Q_\mathrm{int}
\end{equation}

\noindent It can be seen that when $f_0$ increases due to straining the membrane ($f_0 \propto V^2$), the increase in $f_0$ dilutes the intrisic dissipation losses and therefore enhances Q.

\newpage
\section{S4. Device measurements}
\begin{figure}[h]
    \centering
    \includegraphics[width=1.0\textwidth]{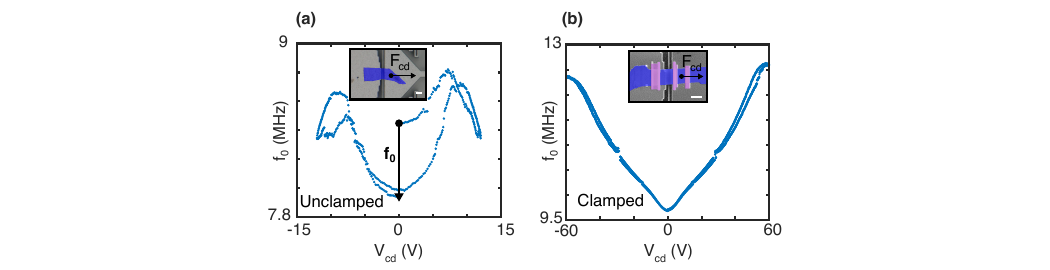}
    \caption{Comparison of the resonance frequency ($f_0$) versus the comb drive voltage ($V_\mathrm{cd}$). $F_\mathrm{cd}$ indicates the pulling direction of the suspended shuttle by the comb drive actuator. Scale bars: 10 micrometers. a) Unclamped device D2, $V_\mathrm{max}$ = 12 V, false-colored optical image (blue: 2D membrane) b) Device D4 clamped with a platinum layer using EBID,$V_\mathrm{max}$ = 60 V (blue), false-colored SEM image (blue: 2D membrane, pink: platinum clamps)}
    \label{fig:f-V-D2-D4}
\end{figure}


\end{document}